\documentclass[a4paper,11pt,times,astrosym]{aastex631}
\usepackage[english]{babel}
\usepackage{amsmath}
\usepackage{graphicx}
\usepackage{natbib}

\submitjournal{ApJ}

\shorttitle{Removing the parameter correlation of ISN gas in direct-sampling observations} 
\newcommand{\kms}{~km~s$^{-1}$}

\newcommand{\cc}{~cm$^{-3}$}

\begin{document}

\title{Breaking correlation in the inflow parameters of interstellar neutral gas in direct-sampling observations}

\correspondingauthor{M. Bzowski}
\email{bzowski@cbk.waw.pl} 

\author[0000-0003-3957-2359]{M. Bzowski}
\affil{Space Research Centre PAS (CBK PAN), Bartycka 18a, 00-716 Warsaw, Poland}

\author[0000-0002-5204-9645]{M.A. Kubiak}
\affil{Space Research Centre PAS (CBK PAN), Bartycka 18a, 00-716 Warsaw, Poland}

\author[0000-0002-2745-6978]{E. M{\"o}bius}
\affil{University of New Hampshire, Durham, NH}

\author[0000-0002-3737-9283]{N.A. Schwadron}
\affil{University of New Hampshire, Durham, NH}

\begin{abstract}
We analyze the reasons for the correlation between the temperature, direction, and speed of the interstellar neutral gas inflow into the heliosphere, obtained in analyzes of observations performed by the IBEX-Lo instrument onboard Interstellar Boundary Explorer (IBEX). We point out that this correlation is the combined result of the inability to measure the speed of the atoms that enter the instrument and the restriction of the observations to a short orbital arc around the Sun performed by the instrument during observation. We demonstrate that without the capability to measure the speed, but with the ability to perform observations along longer orbital arcs, or from at least two distant locations on the orbit around the Sun, it is possible to break the parameter correlation. This, however, requires a capability to adjust the boresight of the instrument relative to the spacecraft rotation axis, such as that of the planned IMAP-Lo camera onboard the Interstellar Mapping and Acceleration Probe (IMAP).
\end{abstract}

\section{Introduction}
\label{sec:intro}
\noindent
The Sun is traversing an interstellar cloud of a partly ionized, magnetized gas. The interaction between this gas and the solar wind is responsible for the creation of the heliosphere \citep{axford_etal:63a}. The ionized component of interstellar matter and the solar wind plasma are separated at the heliopause, and the neutral component penetrates freely into the heliosphere. The flow distribution of this component inside the heliosphere is determined by a combination of solar gravity, ionization losses due to interaction with solar wind particles and solar EUV radiation, and -- for hydrogen -- the solar resonant radiation pressure \citep{patterson_etal:63a}. The main species within the interstellar neutral (ISN) gas include hydrogen and helium, but some of the less abundant components, including oxygen, neon, and deuterium, have also been detected \citep{bochsler_etal:12a, schwadron_etal:16a, rodriguez_etal:13a}. Because of very low densities ($\ll 1$~\cc~ for H and even less for heavier species), the ISN gas throughout the heliosphere can be regarded as collisionless. Observations of ISN gas, its derivative particle components, and the solar light scattered off this gas bring information on the physical state of the matter in the local interstellar medium (LISM). 

The diagnostic potential of ISN gas observations is exploited using three main measurement techniques: (1) observations of the so-called heliospheric backscatter glow, which appears due to the fluorescence of ISN atoms excited by solar emission lines, (2) observations of pickup ions, i.e., a population of ISN atoms ionized inside the heliosphere, subsequently forming a singly charged sub-population in the solar wind, and (3) direct sampling of ISN atoms. Historically first were the discovery observations of the heliospheric backscatter glow of ISN H \citep{morton_purcell:62a, bertaux_blamont:71, thomas_krassa:71}; a review of early observations of the heliospheric glow was presented by \citet{fahr:74}. 

ISN H is strongly depleted inside the heliosphere due to ionization processes and radiation pressure, and its distribution function is strongly modified within the outer heliosheath, i.e., in the region of perturbed  interstellar matter ahead of the heliopause \citep{baranov_etal:91}. Therefore, it is more convenient to use ISN He to infer the Sun's velocity relative to the LISM and the LISM temperature. Helium is abundant in the LISM \citep[the H/He ratio of the neutral components $\sim 12-13$,][]{slavin_frisch:07a, bzowski_etal:19a}, weakly ionized inside the heliosphere \citep[because its ionization rate at 1~au is $\sim 10^{-7}$~s$^{-1}$, compared with $> 5\times 10^{-7}$~s$^{-1}$ for H,][]{rucinski_etal:96a, bzowski_etal:13b, sokol_etal:20a}, and negligibly susceptible to solar radiation pressure. Therefore, at 1~au ISN He is more abundant than ISN H. Furthermore, it is relatively little modified ahead of the heliopause by charge-exchange and elastic collisions \citep{bzowski_etal:17a, swaczyna_etal:21a, fraternale_etal:21a}, which facilitates retrieving the physical state of the unperturbed interstellar medium. 

ISN He has been used for diagnosing the LISM and its interaction with the heliosphere employing all three of the aforementioned measurement techniques. None of them, however, can provide full information on the physical state of the ISN gas -- at least one of the four relevant parameters becomes suppressed. As a result, even though careful analysis yields all parameters, strong correlations in their uncertainties appear. As a result, data are consistent with some combinations of the parameter values much more likely than with the others, which sometimes is referred to as parameter degeneracy or parameter correlation. Visually, such a situation can be described as forming ``tubes'' of likely parameter values in the four-dimensional parameter space.

Direct-sampling measurements performed from 1~au by IBEX \citep{bzowski_etal:12a, mobius_etal:12a, bzowski_etal:14a, wood_etal:15a, bzowski_etal:15a, schwadron_etal:15a, swaczyna_etal:18a,swaczyna_etal:22b} provided the flow direction and speed, and the temperature of the ISN gas with uncertainties strongly correlated with each other \citep{bzowski_etal:12a, mobius_etal:12a, bzowski_etal:15a, mobius_etal:15b, lee_etal:15a, swaczyna_etal:15a}. On the other hand, analyses of PUIs and the helioglow provide consistent conclusions concerning the flow direction of ISN He \citep{vallerga_etal:04a, mobius_etal:15c, taut_etal:18a}, but a systematic difference between estimates of the inflow speed and gas temperature from direct sampling and from helioglow analysis exists, the latter returning a much larger temperature than the former. 

Reducing the uncertainties below $\sim 1\degr${} in the flow direction and $\sim 1$~\kms~in the speed is needed to facilitate studies of ISN H and the secondary population of ISN He. This is because investigation of the secondary population in direct sampling observations requires subtracting the contribution from the unperturbed ISN He population from the measured signal \citep{kubiak_etal:16a, galli_etal:19a, swaczyna_etal:18a}. Also, searching for signatures of hypothetical deviations of the primary ISN gas population from thermal equilibrium, as those manifesting by a kappa distribution function \citep{sokol_etal:15a} or by a temperature anisotropy \citep{wood_etal:19a}, or of the effects of elastic collisions within the outer heliosheath, recently suggested to modify the distribution function of ISN He at the heliopause \citep{swaczyna_etal:21a}, requires a very precise knowledge of the first moments of the distribution function of the ISN gas, i.e., its flow vector. Last but not least, a precise knowledge of the inflow direction of both the primary and the secondary populations of ISN He is needed to determine with a high accuracy the orientation of the approximate symmetry plane of the heliosphere, defined by the flow vector of ISN He and the vector of the local interstellar magnetic field. As shown by \citet{zirnstein_etal:15a} and \citet{kubiak_etal:16a}, the direction of inflow of the secondary population belongs to the plane defined by the aforementioned vectors, which implies that the direction of the so-called B-V plane can be determined by 1 au observations of the primary and the secondary population of ISN He. 

Here, we investigate the reasons for the degeneracies in the ISN He parameters inferred from observations and suggest the observation capabilities required from direct-sampling instruments operating in the vicinity of the Earth's orbit to remove the parameter correlation. We start from presenting the reasons for the correlation of the direction, speed, and temperature of the ISN gas, obtained from direct sampling measurements performed from a spacecraft co-moving with the Earth, like Interstellar Boundary Explorer (IBEX). Using a simple model with the thermal spread of the ISN gas neglected, we approximately reproduce the correlation obtained from the data analysis and we suggest a method to break the parameter correlation (Section \ref{sec:whyDegeneracy}). We then verify the findings using an advanced model of the gas distribution (Section \ref{sec:thermalSpread}), first illustrating the effects of the direction and speed (Section \ref{sec:speedDirCorrel}), and speed and temperature correlation (Section \ref{sec:speedTempCorrel}) for the IBEX viewing conditions, and we proceed to verify the idea of parameter breaking outlined in Section \ref{sec:whyDegeneracy}. We show that the essential prerequisite for the parameter breaking is the ability of the instrument to change its boresight direction and point out that a space mission within the reach of present measurement technology, like the planned Interstellar Mapping and Acceleration Probe (IMAP) mission \citep{mccomas_etal:18b}, will hopefully be able to remove the parameter correlation issue.

\section{Why do the inflow parameters of the ISN gas come out from observations correlated with each other?}
\label{sec:whyDegeneracy}
\noindent
The flow velocity of ISN atoms inside the heliosphere is governed by Sun's gravity. The gravity force bends the straight-line trajectories of the atoms into hyperbolae with the Sun in their foci. Had the ISN gas been infinitely cold \citep[i.e., monoenergetic, without thermal spread, as in the so-called cold model,][]{fahr:68, blum_fahr:70a, axford:72, johnson:72a, johnson:72b, fahr_lay:73a,holzer:77, thomas:78}, at very far distances from the Sun the atoms would move parallel to each other with identical speeds relative to the Sun. As their distances to the Sun drop, the gravity bends their trajectories and deflects them from their original unperturbed direction of motion, depending on the impact parameter of a given trajectory.

A direct-sampling experiment, like IBEX or GAS/Ulysses, detects the atoms in situ while moving around the Sun with a velocity of a magnitude comparable to that of ISN atoms. These instruments have been able to determine the direction from which an atom is coming, but not its impact speed. Had the direction and the speed both been measured, with the orbital velocity of the instrument known, it would be possible to unambiguously determine the velocity vector of the incoming atom in the solar-inertial frame, and subsequently the velocity of motion of this atom far away ahead of the Sun. This velocity would be equivalent to the inflow velocity of ISN gas.
\begin{figure}
\centering
\includegraphics[width=0.5\textwidth]{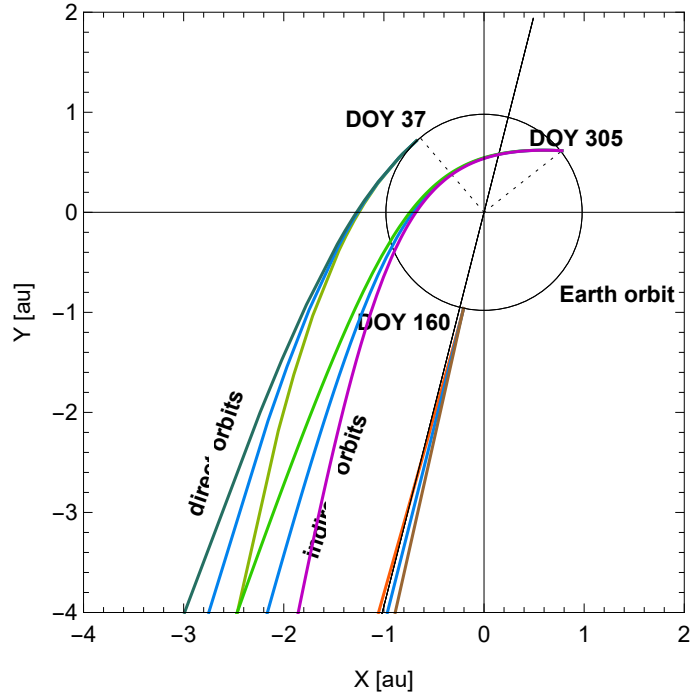}
\caption{
\emph{degenOrbits} 
Example plots of orbits that impact at an instrument moving with the Earth from identical directions but with different speeds. The black circle represents the Earth orbit, the Earth is moving counterclockwise. When the instrument is at the locations indicated by the dotted radius-vectors and the DOY labels, it is hit by three atoms approaching from the same direction but with different speeds. The plot is drawn assuming that one of these atoms has the velocity vector far away from the Sun identical to the inflow velocity of ISN gas. The other two orbits impact the instrument with a speed greater (black) or smaller (green) by 2 \kms{} than that of the former atom. Their orbits, and thus the directions of motion ahead of the Sun clearly differ. The observer location for DOY 37 is selected so that it corresponds to the viewing geometry of IBEX-Lo, with the atoms impacting the instrument at the directions perpendicular to the radius vector. The blue, orange, and brown lines mark detection geometry at an alternative location, when the instrument is at the upwind longitude (DOY 160); viewing of these atoms is impossible for IBEX. For these two locations, direct orbits are shown. For DOY 305, three indirect orbits are presented; these are also inaccessible for IBEX, but will be accessible for IMAP.}
\label{fig:degenOrbits}
\end{figure}

However, since the impact speed is not determined, a family of solutions exists that feature different impact speeds and identical directions of impact at a moving instrument. Such a situation is illustrated in Figure \ref{fig:degenOrbits}, where we show two families of orbits, reaching the instrument on the spacecraft traveling in Earth orbit. While in the instrument-inertial frame the directions are identical, the magnitudes of the speed differ, which results in different velocity vectors of the atoms far away from the Sun. Such families of orbits, related by the impact geometry at the spacecraft, exist for all locations along the instrument orbit around the Sun. As a result, an observer who performs the observations during a given day of the year (DOY) will find a degeneracy of the direction and speed of inflow of the ISN gas. These degenerate directions and speeds of the atoms at infinity form lines in the parameter space.

This effect can be simulated as follows. Adopting a velocity vector of the ISN gas at infinity and selecting a location of the instrument at the Earth's orbit, one can calculate the relative velocity vector of the atom. With this, one can vary the impact speed by a certain value, transform the varied impact velocity vectors into the solar-inertial frame, and subsequently, by solving the hyperbolic Kepler equation, find the corresponding velocity vectors at infinity. Using this recipe, one can obtain projections of the correlation tubes on the longitude-speed plane for any selected DOY. We performed such calculations adopting, after \citet{bzowski_etal:15a}, the inflow speed 25.764~\kms, the flow longitude 75.75\degr, and the flow latitude $-5.16\degr$. The magnitude of the speed variation at the instrument was adopted as $\pm 2$ and $\pm 4$~\kms{} in the spacecraft-inertial frame, based on the insight from simulations performed using the Warsaw Test Particle Model for ISN He \citep[nWTPM, ][]{sokol_etal:15b}, which showed that the thermal speed parallel to the ISN flow in the spacecraft frame at 1 au, calculated as the second central moment of the atom speed distribution at the instrument in the spacecraft-inertial frame, is equal to $\sim 2 \pm 0.5$~\kms. This is adopted as the spread of relative speeds along the impact direction at the instrument.

\begin{figure}
\centering
\includegraphics[width=0.4\textwidth]{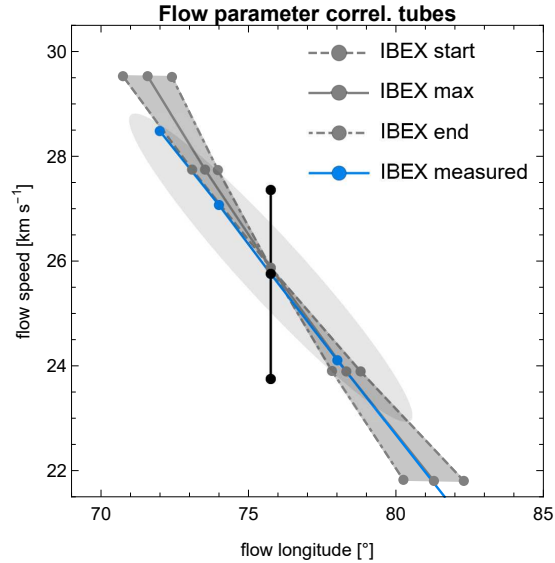}
\caption{
{\emph{vlCorrelTubesIBEX}} 
Simulated and actual correlation tube for the ISN He observation conditions for IBEX-Lo. A projection of the tube in the longitude--speed plane in the four-dimensional parameter space is shown. The gray-shaded region bound by the dashed and dash-dotted lines marks the simulated correlation lines for DOYs 22 and 57, which represent the typical time interval of ISN He observations by IBEX-Lo each year. The maximum of the observed flux occurs at DOY 37 (gray solid line, IBEX max). The gray ellipse represents the uncertainty of the flow parameters typically obtained for IBEX-Lo observations collected during one observation season (a 3-$\sigma$ uncertainty). The blue line represents the correlation tube actually obtained from analysis of IBEX observations by \citet{bzowski_etal:15a}. The gray dots represent the speed and ecliptic longitude from the correlated parameter sets listed in the first two rows in Table \ref{tab:degenParam} (the speed--direction correlation). The black dots illustrate a projection on the (speed--longitude) parameter space subspace of the parameter sets from the last two rows in Table \ref{tab:degenParam}, corresponding to the parameter correlation by the Mach number. Clearly, the latter two points are selected to be far away from the correlation tube for the speed--direction correlation tube. }
\label{fig:vlCorrelTubesIBEX}
\end{figure}

\begin{figure}
\centering
\includegraphics[width=0.485\textwidth]{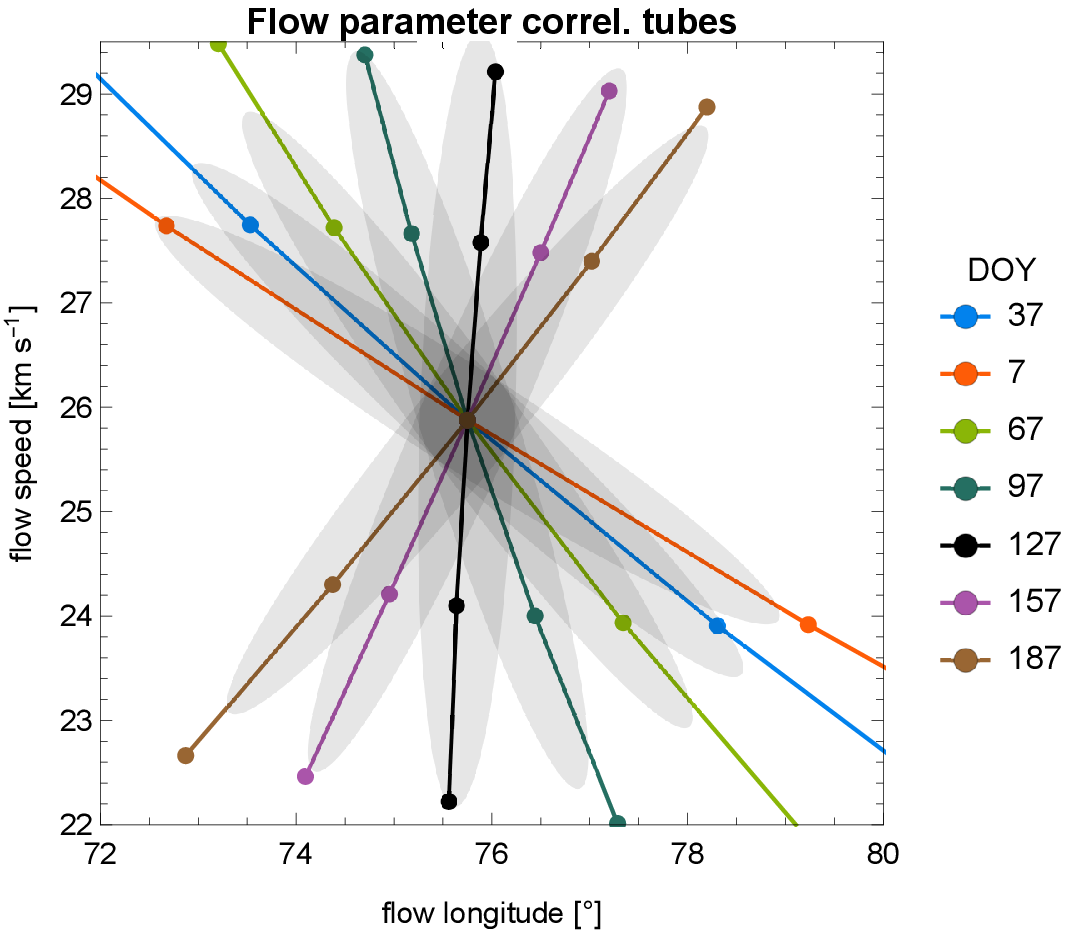}
\includegraphics[width=0.420\textwidth]{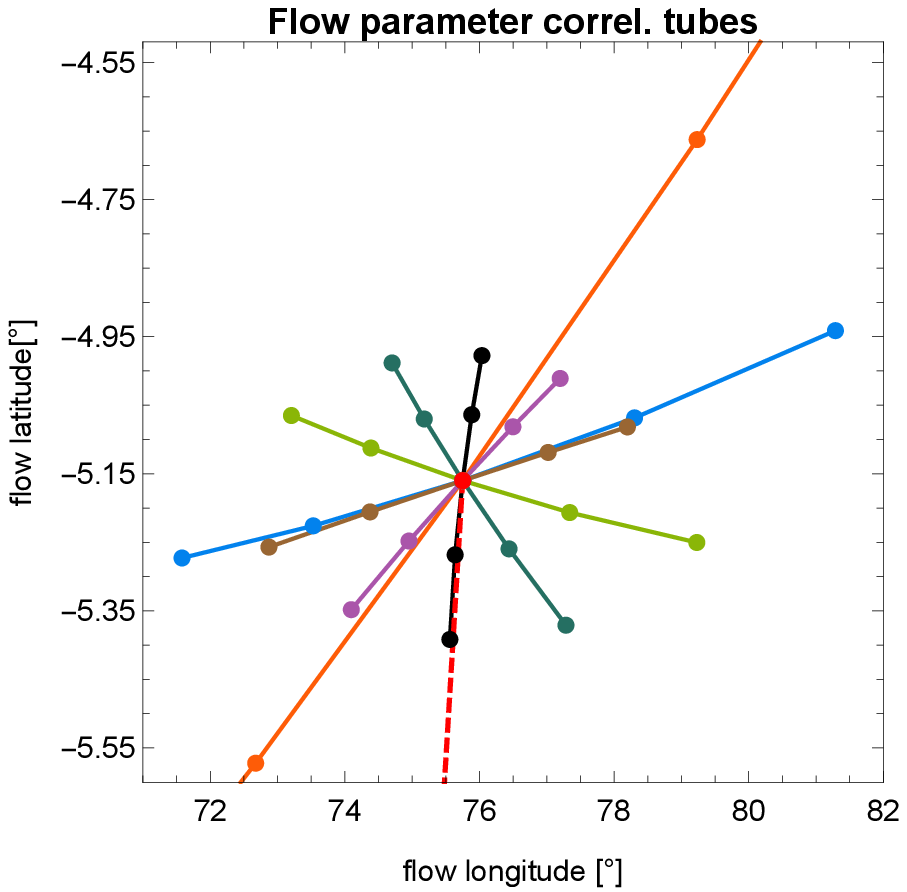}
\caption{
\emph{CorrelTubesIMAP} 
Left panel: Simulated correlation tubes projected on the longitude-speed plane for uniformly spaced DOYs during the first half of calendar year. The IBEX max correlation line, copied from Figure~\ref{fig:vlCorrelTubesIBEX}, is drawn in blue. The gray ellipses are cartoon illustrations of the correlation tubes for individual correlation lines. The intersection of these regions, visible as the dark region near the center, illustrates the reduction of the parameter uncertainties, expected when observations taken during several well-separated time intervals during the year are available. Right panel: correlation tubes projected on the longitude-latitude plane, with identical color coding to that in the left panel. Additionally, the broken line, almost coincident with the line for DOY 127, connects the longitude and latitude of the flow directions of the primary ISN He and the Warm Breeze, i.e., the secondary population of interstellar He.}
\label{fig:CorrelTubesIMAP}
\end{figure}

Formation of a correlation tube is presented in Figure~\ref{fig:vlCorrelTubesIBEX} for the IBEX-Lo observation interval. The optical axis of the IBEX-Lo camera is perpendicular to the rotation axis of the IBEX spacecraft, which is maintained approximately pointing at the Sun \citep{fuselier_etal:09b, mobius_etal:09a}. Following \citet{bzowski_etal:15a} and \citet{swaczyna_etal:18a}, we select a time interval during the year from which IBEX-Lo observations of ISN He were taken to the analysis by these authors, i.e., DOYs 22 and 57, and additionally the DOY for the maximum of the observed signal, i.e., DOY 37. We calculate the longitude--speed correlation lines for the start, maximum, and end days of this interval and plot the resulting correlation tube in Figure~\ref{fig:vlCorrelTubesIBEX} as the gray region. Superimposed, we plot the actually obtained correlation line from the data analysis by \citet{bzowski_etal:15a} (see their Figures 5 and 7). Clearly, the observation-based correlation line overlaps with the simulated tube, even though the simulated tube was obtained using very simplified assumptions. 

This part of the analysis suggests that performing direct-sampling observations of ISN He along a relatively short arc of the Earth's orbit around the Sun (about 35\degr, as in the case of IBEX-Lo observations) will not permit to fully remove the correlation of the ISN flow parameters. The parameter correlation exists because of ballistic reasons in the situation when one can measure the direction of inflow of the gas at the instrument, but not its speed. A detailed analysis, like that proposed by \citet{swaczyna_etal:15a} and used by \citet{bzowski_etal:15a}, \citet{swaczyna_etal:18a} and \citet{swaczyna_etal:22b}, enables determining the most likely solution, but with uncertainties given by a complex covariance matrix, which results in the largest uncertainties along the correlation lines. Therefore, increasing the counting statistics or reducing the background will not be sufficient to remove the parameter correlation. The  correlation may be partly alleviated by varying the spin axis of the spacecraft around the ecliptic plane, as pointed out by \citet{schwadron_etal:22a}. However, performing observations over a long orbital arc, or during at least two time intervals along the orbit separated in time by several months, will result in the creation of several correlation tubes intersecting at large angles in the parameter space near the true inflow parameters, and thus will constrain the parameters more tightly.
\begin{figure}
\centering
\includegraphics[width = 0.4\textwidth]{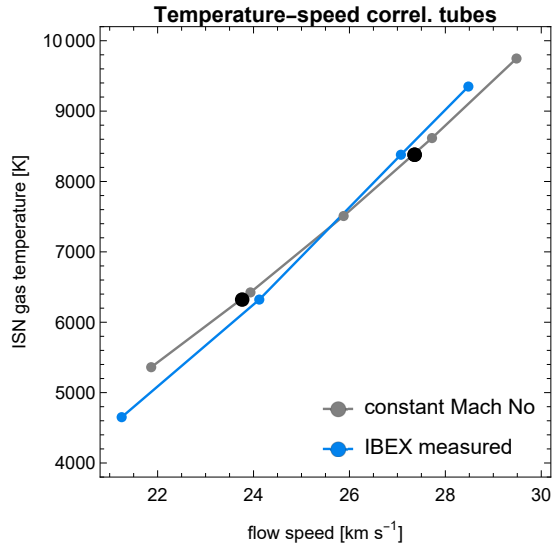}
\caption{
{\emph {TvCorrelTubesIBEX }}
Temperature--speed correlation lines obtained from IBEX-Lo observations (blue) and simulated using the speed values from the simulated longitude-speed correlation line for DOY 37 (see Figure~\ref{fig:vlCorrelTubesIBEX}) and the Mach number as obtained from IBEX-Lo observations using Equation~\ref{eq:TfromMach} (gray). The observed Mach number is equal to 5.075. The thick black dots represent the temperature--speed pairs of the parameter sets correlated by the Mach number, listed in the last two rows in Table \ref{tab:degenParam}.}
\label{fig:TvCorrelTubesIBEX}
\end{figure}

This is illustrated in the left panel of Figure~\ref{fig:CorrelTubesIMAP}, which shows correlation tubes projected on the longitude-speed plane, simulated for DOYs uniformly distributed along the first half of the year. The correlation lines rotate around the point defined by the inflow parameters of ISN He assumed in the simulations. Direct-sampling observations performed during a few consecutive  days during any portion of the year will result in parameters correlated along a correlation tube specific for the given observation time. Cartoons of such tubes are plotted as the elliptical gray shades. The lengths of the axes of these elliptical region were adopted based on the approxmation provided by \citet{swaczyna_etal:15a}. Inspection of the correlation lines and the associated parameter tue regions represented by the elliptical shades shows that by combining observations from at least two time intervals of $\sim 30$ days separated by $\sim 60$ days is expected to constrain the parameters so that the correlation is mostly removed. The resulting uncertainty range of the parameters is represented by the intersections of the parameter tubes, illustrated by the dark regions. Clearly, if observations from long orbital arcs are available, then the uncertainty becomes close to spherical in the parameter space, and the size of the uncertainty region is reduced by a large factor.

This is also true for the correlation between the longitude and latitude of the inflow direction, as show in the right panel of Figure \ref{fig:CorrelTubesIMAP}.

In the discussion above, the finite temperature of the ISN gas has been neglected. However, analysis of IBEX-Lo observations showed that the temperature is correlated with speed and inflow direction, forming a parameter tube. It was also found that the Mach number for these correlated parameters is practically independent of these parameters. The reason for this is the thermal spread of atom velocities. The differentiation of atom speeds at the instrument exists because the thermal spread of the ISN translates into the observed angular distribution of the flow. Figure~\ref{fig:TvCorrelTubesIBEX} demonstrates this clearly. We compare the IBEX correlation line, drawn  in blue, with a relation obtained in the following way: take the speeds from the longitude-speed correlation line presented in Figure~\ref{fig:TvCorrelTubesIBEX} and calculate the corresponding temperatures assuming a constant Mach number using the formula: 
\begin{equation}
T = \frac{3 m_{\text{He}}}{5 k_{\text{B}}}\left(\frac{v}{M} \right)^2,
\label{eq:TfromMach}
\end{equation}
where $M$ is the Mach number, $v$ the speed of the ISN gas in the unperturbed interstellar medium, $k_\text{B}$ the Boltzmann constant, and $m_{\text{He}}$ the mass of He atoms. With this, the temperature-speed parameter tube is identical for all DOYs. Inspection of Figure~\ref{fig:TvCorrelTubesIBEX} shows that the observation and model correlation lines agree very well with each other. The flow longitude--temperature correlation exists because the temperature is correlated with the speed, and the speed with the longitude. Hence, constraining the temperature is obtained when the longitude--speed correlation is removed. 

An interesting aspect of the parameter correlation is pointed out for DOYs about 121 (the black line in Figure \ref{fig:CorrelTubesIMAP}). The correlation line is vertical in the figure, which implies that observations for this DOY will return an uncertainty in speed, but relatively little in the inflow direction. That means that this DOY seems advantageous for precise determination of the inflow direction. 

The discussion presented so far is mostly based on simple-model arguments. However, the existence of intersecting parameter correlation tubes and the feasibility of using them to constrain the ISN flow vector and gas temperature have been verified in an analysis of ISN He observations performed by the Ulysses/GAS experiment \citep{bzowski_etal:14a}. The GAS experiment \citep{witte_etal:92a, witte_etal:93} was the first to directly sample ISN He. The data were collected along long arcs of the Ulysses orbit around the Sun, at the perihelion side of the orbit between the north and south solar polar directions \citep[see Figure 1 in][]{bzowski_etal:14a}. The spacecraft was three axis-stabilized, and the beam of ISN He was maintained within the field of view by frequent adjustments of the direction of the instrument boresight. Ulysses performed three revolutions around the Sun, and usable data were split into several groups \citep[see Figure 14 in][]{bzowski_etal:14a}. Determination of the ISN He flow parameters performed by these authors using data from these individual arcs resulted in the discovery of intersecting parameter tubes in the parameter space (see Figures 11 and 12 in this paper). These intersections were used to constrain the flow parameters. Simultaneously, these authors pointed out that the correlation tubes obtained from similar arcs in different Ulysses orbits are very similar to each other (see Figures 8 and 9 in the aforementioned paper). This is, of course, expected based on the insight presented earlier in our paper. Similarly, \citet{wood_etal:15a} obtained well-constrained inflow parameters based on Ulysses observations collected on a long orbital arc, even though they used a different parameter determination method to that used by \citet{bzowski_etal:14a}. 

\section{Effects of the thermal spread of atom velocities}
\label{sec:thermalSpread}
\noindent
In this section, we acknowledge that the ISN gas has a thermal spread. We demonstrate by simulations of the ISN He signal performed using a state of the art hot model of ISN He (WTPM) for the viewing conditions of IBEX-Lo and IMAP-Lo that indeed, the simulated signals obtained for the parameters highly-correlated based on the insight from the cold model,  are indeed almost indistinguishable in the relevant portions of the Earth orbit, in other words, these parameters are indeed degenerate.  However, the signals calculated for the same parameter sets but for different portions of the Earth's orbit (i.e., for different instrument location) become to different between the different parameter sets. Thus, the degeneracy obtained for the IBEX viewing conditions can be removed by performing observations in a different portion of the Earth orbit. This, however, is only feasible with the capability to shift the boresight of the instrument, which will be available on IMAP-Lo. We also demonstrate that the signals expected for uncorrelated parameter sets for the IBEX viewing geometry become hardly distinguishable for the IMAP-Lo viewing geometry at the locations in the Earth's orbit predicted by the theory presented in the previous section, which illustrates that the reasons for the parameter degeneracy are well understood.

Subsequently, we demonstrate that a similar behavior of the simulated signal, and thus the parameter correlation, is expected not only for the boresight orientation optimized for the maximum statistics, but also for alternative viewing conditions provided that a portion of the thermally-broadened ISN beam is in the instrument field of view. In fact, observing flanks of the distributions instead of the peaks may facilitate removing the inflow parameter correlation at the cost of a certain reduction in counting statistics. 

We start from the removal of the direction--speed correlation, followed by the temperature--speed or temperature--direction correlation.

\subsection{Simulations and parameter selection}
\label{sec:paramSel}
\noindent
To investigate the parameter correlation removal, we performed a series of numerical simulations of the signal due to ISN He expected to be observed by an IMAP-Lo-like instrument orbiting the Sun at 1 au in the ecliptic plane. The virtual instrument is supposed to be mounted on a virtual spin-stabilized spacecraft with the spin axis directed precisely at the Sun for each DOY, and the elongation angle of the boresight of this instrument from the spin axis can be freely adjusted. The change of the  ecliptic longitude of the spacecraft during a day was neglected.
\begin{deluxetable*}{llllll}[!h]
\tablecaption{\label{tab:degenParam} degenParam Flow parameters of ISN He used in the simulations}
\tablehead{case & longitude [\degr] & latitude [\degr] &  speed [\kms] & temperature [K] & Mach no}
\startdata
nominal set    & 75.75 & $-5.160$ & 25.784 & 7443 & 5.0754 \\
degen. 1       & 74.00 & $-5.287$ & 27.073 & 8380 & 5.0262 \\
degen. 2       & 78.00 & $-4.995$ & 24.120 & 6322 & 5.1556 \\
\hline
const. Mach 1  & 75.75 & $-5.160$ & 23.763 & 6322 & 5.0754 \\
const. Mach 2  & 75.75 & $-5.160$ & 27.359 & 8380 & 5.0754 
\enddata
\end{deluxetable*}

The simulations were performed assuming the inflow parameters of the ISN gas identical to those obtained from analysis of IBEX observations and the two alternative sets of highly correlated parameters, listed in the first three rows in Table \ref{tab:degenParam}. The differences between the optimum parameters, listed as ``the nominal set'', and the two alternative sets, listed as ``degen. 1'' and ``degen. 2'', are chosen so that the parameters lie on the correlation tube in the parameter space. This tube is shown with the blue line in Figure \ref{fig:vlCorrelTubesIBEX}. The speed--longitude combinations of the two latter sets are drawn as purple dots.

Another series of simulations was performed for two sets of inflow parameters that are located outside the parameter tube. These parameter sets are correlated with each other and with the nominal parameter set (line 1 in Table \ref{tab:degenParam}) by having identical Mach number. These two parameter sets are listed in the last two rows in Table \ref{tab:degenParam} as ``const. Mach 1'' and ``const. Mach 2''. They are plotted with the blue line in Figure~\ref{fig:TvCorrelTubesIBEX} and as thick black dots in Figure \ref{fig:vlCorrelTubesIBEX}. In this case, the inflow direction (longitude and latitude) was identical between the three parameter sets, and the speeds and temperatures were varied so that they correspond to the Mach number corresponding to the nominal parameter set.

All simulations were carried out using a time- and heliolatitude-dependent model of the ionization factors, adopted from \citet{sokol_etal:19a}. For the virtual observation year, we chose 2015 to simulate the conditions of high solar activity, resulting in relatively high ionization losses of ISN He inside the heliosphere. This choice was made for correspondence with the expected level of solar activity after launch of the IMAP mission in 2025. 

The simulations were performed using the numerical version of the Warsaw Test Particle Model \citep[nWTPM; ][]{sokol_etal:15b}. They modeled the flux of ISN He filtered by an IMAP-like collimator as a function of the spacecraft spin angle. More details can be found in \citet{sokol_etal:19c}. The results shown further in the paper are organized into series of several selected DOYs. The flux shown is normalized by the maximum flux for the presented simulation series for a given parameter set and the selection of DOYs and elongation angles. Separate normalization constants are calculated for series of simulations with different flow parameters. This normalization is performed to eliminate differences in the absolute flux between the cases. This is needed because in reality, the absolute sensitivity of an instrument is known with a limited accuracy, so determination of the ISN gas parameters cannot rely on the absolute calibration of the instrument. In fact, we only compare the shapes of the simulated flux as a function of the spacecraft spin angle. 

For the presentation of the results, we selected a relatively high cutoff of 1\% of the maximum flux, to approximately account for the expected presence of the secondary population of ISN He, the Warm Breeze, as pointed out by \citet{sokol_etal:19c}. We also used an energy sensitivity limit, adopted at 20 eV for all discussed species, following the insight from \citet{galli_etal:15a,sokol_etal:15a}.

\subsection{Removing the direction--speed correlation}
\label{sec:speedDirCorrel}
\subsubsection{IBEX viewing conditions: parameter degeneracy or correlation?}
\label{sec:ibexViewing}
\noindent
The Interstellar Boundary Explorer (IBEX) is the first spacecraft to sample ISN gas directly at the Earth's orbit \citep{mccomas_etal:09a}. The spin-stabilized spacecraft is in an elongated orbit around the Earth \citep{mccomas_etal:11a} with a period of several days. The IBEX-Lo instrument, used to observe the ISN gas \citep{fuselier_etal:09b, mobius_etal:09a}, has a boresight with a fixed direction, inclined at 90\degr{} to the spacecraft spin axis. The spin axis is shifted once or twice per IBEX orbit (i.e., every several days) to approximately follow the Sun. Thus, between the spin axis shifts, the beam of the ISN gas viewed by the instrument is slowly moving across the observed strip in the sky. Because of the fixed viewing direction, the beam of ISN gas can be observed only during several weeks each year. The ISN observation season is further limited by strong contributions to the signal from the secondary population of ISN He at the beginning of each year \citep{bzowski_etal:12a, kubiak_etal:14a, kubiak_etal:16a} and from ISN H later during the year \citep{saul_etal:12a, schwadron_etal:13a,katushkina_etal:15b, galli_etal:19a, rahmanifard_etal:19a}. Effectively, the observation season of the primary population of ISN He spans approximately DOYs 22--57 each year. Thus, the corresponding Earth's orbital arc is $\sim 35\degr$, which makes it short in comparison with a quarter of the Earth orbit that is best suited for breaking the degeneracy (cf the correlation lines for DOYs 37 and 127 in Figure \ref{fig:CorrelTubesIMAP}).

Several analyses of IBEX-Lo observations returned the inflow velocity vector and the temperature of ISN He with uncertainties forming a ``tube'' in the parameter space  \citep{bzowski_etal:12a, mobius_etal:12a, bzowski_etal:15a, schwadron_etal:15a, swaczyna_etal:18a, swaczyna_etal:22b}. The uncertainties of the parameters are much larger along these correlation lines in the four-dimensional parameter space, but much tighter constrained across them. This has been referred to as parameter correlation or parameter degeneracy \citep{schwadron_etal:22a}. 

\begin{figure}[!h]
\centering
\includegraphics[width=1.0\textwidth]{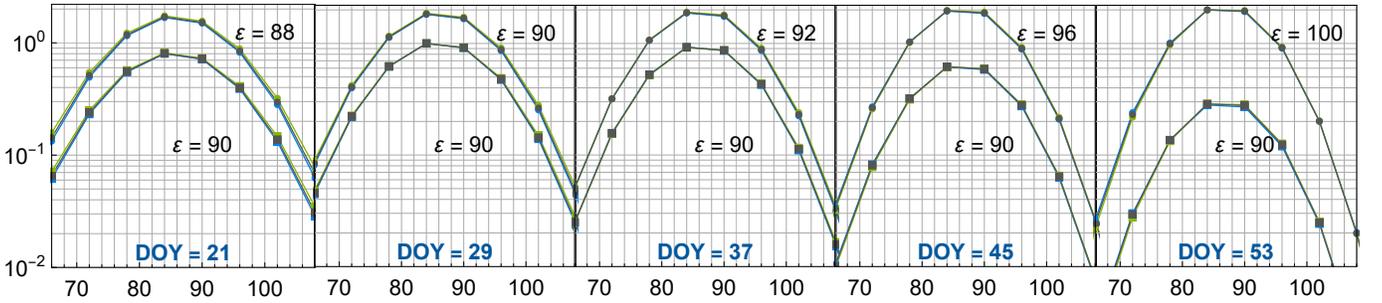}
\caption{\emph{ibexGeom} Two series of simulated normalized fluxes of ISN He for selected DOYs, corresponding to the IBEX ISN observation season: (1) for the boresight adjusted daily to follow the peak of ISN He (dots, upper set of lines) and (2) simulated for an IBEX-like viewing geometry (squares, bottom set of lines), i.e., with the boresight inclined at 90\degr{} to the spin axis $(\varepsilon = 90)$. Three cases of the degenerate inflow parameters taken from Table \ref{tab:degenParam} are drawn: the baseline (black), degen 1 (blue), and degen 2 (green). The horizontal axis represents the spacecraft spin angle. The vertical axis is dimensionless. The variable-elongation case is scaled upward by a factor of 2 to improve the visibility. Discerning the three lines for a given $\varepsilon$ requires a strong magnification of the figure because the normalized fluxes are almost identical between the three degenerate parameter sets. Note that the two line sets drawn for DOY 29 are identical. The DOYs of the simulations are listed in the blue color at the bottom of the panels, and the elongation angles of the boresight from the spacecraft spin axis $\varepsilon$ used for the two groups of simulations in each panels are listed at the corresponding line groups.} 
\label{fig:ibexGeom}
\end{figure}

Discussion provided in Section \ref{sec:whyDegeneracy} suggests that the parameters of the ISN gas determined from an instantaneous observation will be degenerate if the speed of the incoming atoms cannot be measured. If the measurements are performed during a number of consecutive days, then the parameter degeneracy line will be slowly rotating in the parameter space. The simulations discussed in Section \ref{sec:whyDegeneracy} were performed for individual atoms and assuming an idealized situation when the instrument is able to look precisely into the beam of the incoming atoms. However, the actual observation conditions on IBEX differ from these ideal ones, and the sampling was performed along a finite arc of the Earth orbit. As a result, the parameters obtained from the analysis are strongly correlated with each other, but not degenerate, since it was possible to obtain a unique set of the parameters by means of chi-squared analysis \citep[e.g.,][]{swaczyna_etal:15a}. 

To show that the parameter degeneracy or correlation is indeed supported by models of the ISN He flux observed at 1 au, we performed simulations of the IBEX signal for two alternative observation scenarios: (1) for an idealized situation, when the instrument has a variable elongation of its boresight from the spacecraft spin axis, with the boresight always directed so that the peak of the ISN beam goes into the instrument field of view (``follow the peak''), and (2) for a more realistic situation for IBEX, when the boresight is always perpendicular to the spin axis (``stepwise adjustment''). In both cases, the spin axis was assumed to point precisely towards the Sun for each DOY, unlike in the reality on IBEX. The simulations were performed for selected five DOYs during the yearly ISN observation season; covering the time interval chosen by \citet{bzowski_etal:15a}, and subsequently \citet{swaczyna_etal:18a} and \citet{swaczyna_etal:22b} for analysis of the actual observations. Also the selection of spin angle boundaries follows the choices made by these authors. The evolution of the elongation from the Sun of the ISN gas beam during the year for scenario (1) was adopted from \citet{sokol_etal:19c}.

The results are presented in Figure \ref{fig:ibexGeom}. The figure compares the flux for the nominal set of inflow parameters (the black lines) and for the parameters listed in the second and third rows in Table \ref{tab:degenParam}. 

 The signal for scenario (1) is represented by the upper groups of lines, which were scaled by a facor of 2 for a better visual separation from the other set of lines. For the viewing conditions (1), the three simulated cases are indeed degenerate in the sense that the simulations return practically identical results for the three different parameter sets. This illustrates that simulations performed taking the thermal spread of the ISN gas into account confirm the presence of a high parameter correlation obtained from observations performed along a short arc around the Sun, when the instrument boresight is adjusted to exactly follow the peak of the ISN gas beam. Note, however, that the fluxes for the three parameter sets are not perfectly equal. We verified that the differences between the three cases are within $\sim 10$\% and are the largest for the earliest and the latest day in the presented sample. 

In reality, the IBEX-Lo boresight is fixed at 90\degr. The simulations performed for the same DOYs and the same parameter sets are presented as the lower sets of lines in Figure \ref{fig:ibexGeom} (scenario 2). Clearly, also in this case, the three strongly correlated parameter sets return very similar, albeit not identical fluxes. Details of the differences between the three simulated cases are different, but generally their magnitudes are similar to those obtained in scenario (1). This illustrates that the thermal character of the gas allowed to fit a unique solution to IBEX-Lo observations, even though the parameters of the ISN gas flow  were obtained strongly correlated. 

A conclusion from this portion of our study is that indeed, where the simple reasoning presented in Section \ref{sec:whyDegeneracy} suggests a parameter degeneracy, the models of the signal obtained from our realistic simulations predict almost identical behaviors of the simulated flux for the correlated parameter sets. Hence, the parameter degeneracy for these short arcs is confirmed by simulation. 

\subsubsection{Breaking the parameter correlation}
\label{sec:correlBreak}
\noindent
In this section, we will verify if performing observations along longer orbital arcs of the Earth facilitates breaking the parameter correlation and what scenarios for the boresight angle adjustments are the most advantageous to accomplish this goal. Clearly, the differences between the correlated cases should be as large as possible, but on the other hand, the absolute magnitude of the expected signals must be sufficiently large to provide sufficient counting statistics.
\begin{figure}
\includegraphics[width=1.0\textwidth]{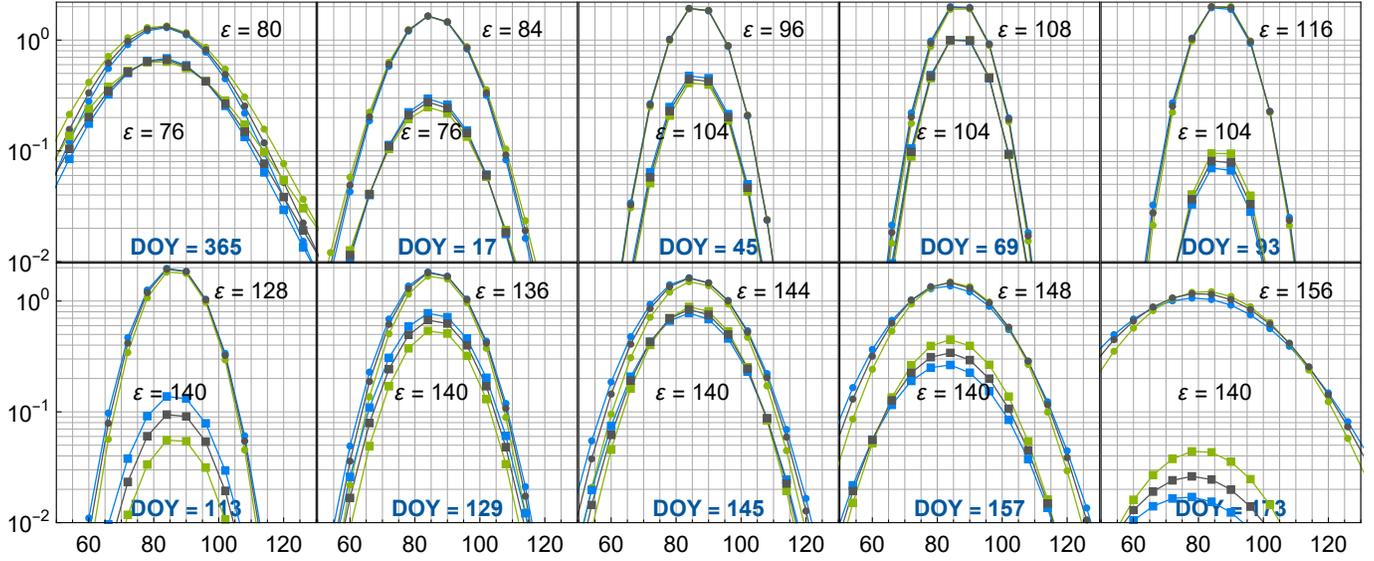}
\caption{{\emph{followVsStep}} Two series of the normalized flux of ISN He for selected DOYs, illustrating the feasibility of breaking the correlation of the ISN flow parameters. The capability of adjusting the elongation angle of the instrument boresight to the spacecraft spin axis is used. (1) dots: The elongation angle set always follows the peak of the primary population of ISN gas, (2) squares: The elongation is changed stepwise with time and maintained constant during prolonged intervals. The color code and the format of the figure are identical with those used in Figure \ref{fig:ibexGeom}. (1) is scaled upward by a factor of 2 to improve the visibility. The DOYs and elongations are listed in the panels.}
\label{fig:followVsStep}
\end{figure}

To that end, we performed simulations of the signal along the Earth orbit throughout the entire year every four days for the three correlated parameter sets, assuming either that the elongation angle is adjusted daily so that the peak of the flow of the ISN gas is within the field of view, or that this angle is adjusted stepwise, with prolonged intervals of a constant setting. The most salient results are presented in Figure \ref{fig:followVsStep}, which has a format similar to that adopted for Figure \ref{fig:ibexGeom}. 

It is clearly visible that an opportunity to break the parameter correlation is obtained by extending observations along the Earth's orbital arc. Within scenario (1) (``follow the peak''), very similar signal shapes are visible for the times of the IBEX observations, but for other DOYs the differences are larger. For observations at the turn of the years (represented by the panel for DOY 365) and several months later, about DOYs 157 and 173, the signals for the selected parameter sets are very different from each other. This means that the parameters that came out correlated during the IBEX observation times will come out uncorrelated from observations performed during different epochs, exactly as suggested by the analysis presented in Section \ref{sec:whyDegeneracy}.

The simulations presented in Figure \ref{fig:followVsStep} were made for parameter sets that are within the correlation lines characteristic for early DOYs during the year (see Figures \ref{fig:vlCorrelTubesIBEX} and \ref{fig:CorrelTubesIMAP}). However, a qualitatively similar behavior is expected for all other (longitude--speed) pairs. For such pairs, the expected signal will be almost identical during some portion of the year, but very different for a different portion, facilitating breaking the parameter correlation provided that observations are performed along a sufficiently long arc along the orbit around the Sun.

It is worth to point out in the context of the planned observations by IMAP-Lo that the effect of breaking the parameter correlation can be obtained also when using alternative scenarios for modification of the elongation angle. This is illustrated by the example presented as the second (lower) set of lines in the panels of Figure \ref{fig:followVsStep}. Using the elongations listed in the figure, the peak of the ISN beam is not in the field of view and the instrument is looking at the flanks of the beam except for one of the orbits. The elongation angle is maintained fixed during prolonged intervals and changed from time to time. This corresponds to the ``stepwise scenario'' (2). Clearly, in this scenario, the signal is weaker than in the former case, but the differences between the signals obtained for the three correlated cases are much larger. However, when planning the observations, one should take the expected statistics into account. Weaker signals imply a larger Poisson noise in the data and large differences, as those seen in lower set of lines the lower-right panel, might be partially obscured by the statistical scatter. Nevertheless, the potential to break the parameter correlation remains. 

When planning the observations, one needs to take several aspects into account. One of them is the presence of other components of the ISN gas in the considered field of view. In the case of ISN He, it is the presence of the secondary population of this species, i.e., the Warm Breeze. This aspect was illustrated in Figure 8 in \citet{sokol_etal:19c}. In the examples presented in this paper, we verified by simulations (not shown) that the expected signal from the Warm Breeze is low relative to that from the primary population of ISN He (around the lower bound of the plots). 

\subsubsection{Using the indirect beam of the ISN gas for breaking the parameter correlation}
\label{sec:ndirBeam}
\noindent
An interesting option for breaking the parameter correlation may be using the indirect beams of the ISN gas. The cold model of the ISN gas inside the heliosphere predicts that in any point in space, two co-planar orbits of ISN atoms intersect, with different impact parameters and the angular momentum vectors oriented oppositely relative to the orbital plane. One of them is referred to as the direct orbit, and the other one as the indirect orbit. The latter one has typically already passed its perihelion and is receding from the Sun, with a positive radial velocity. An example of indirect orbits is presented in Figure \ref{fig:degenOrbits} (DOY 305). Since in reality the ISN gas has a thermal spread, we speak of the direct and indirect beams of ISN atoms. 
 
\begin{figure}
\centering
\includegraphics[scale=0.8]{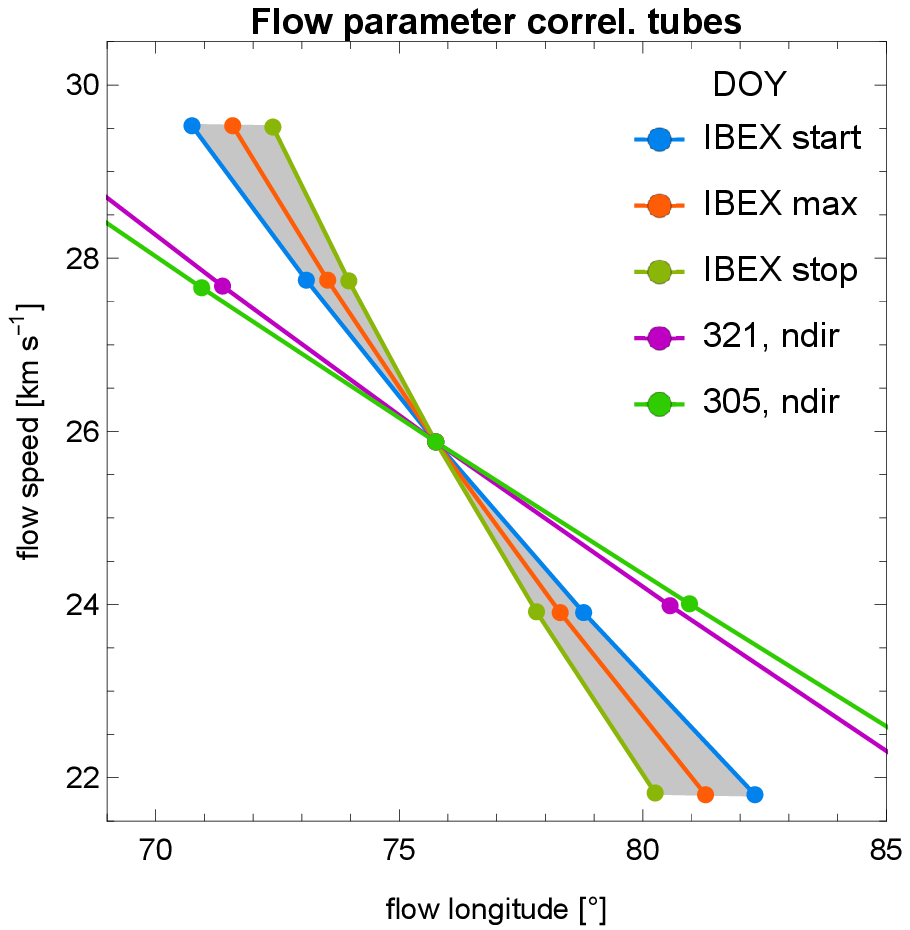}
\includegraphics[scale=0.8]{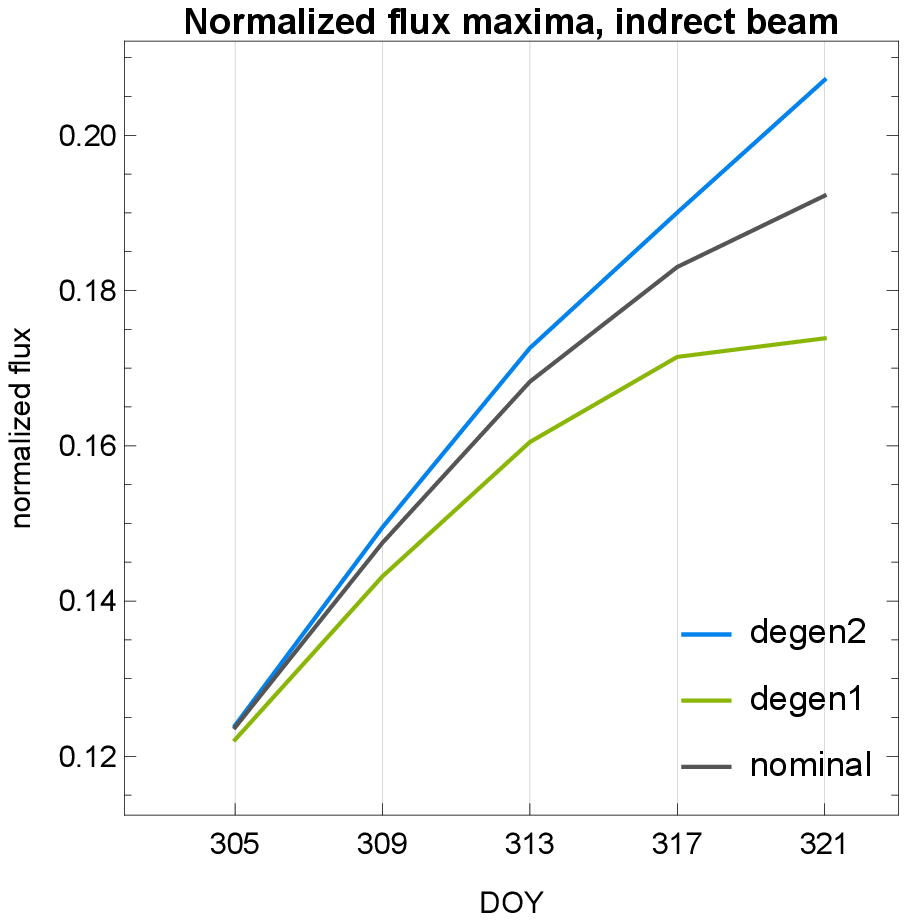}

\includegraphics[width=1.0\textwidth]{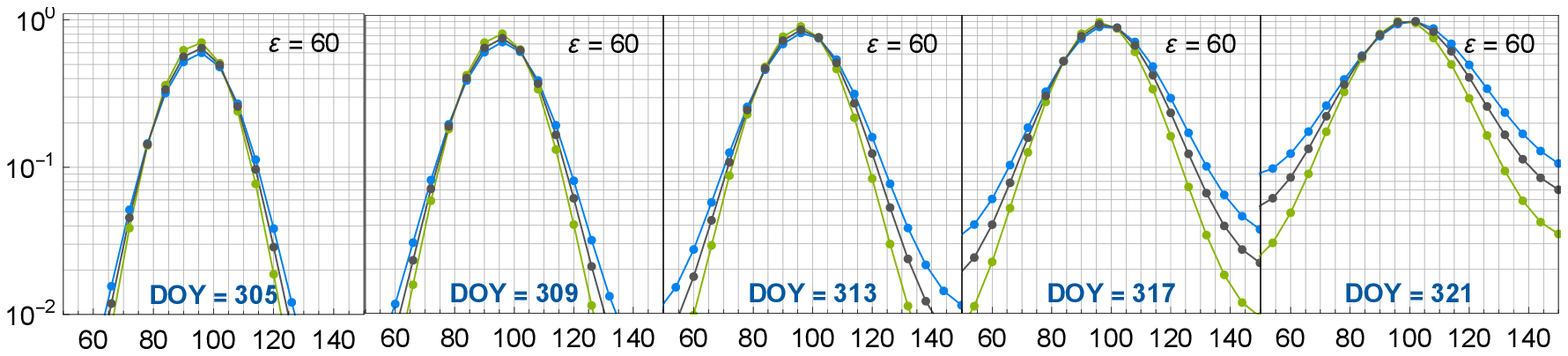}
\caption{
{\emph{NdirCorrelTubes}}
Breaking the parameter correlation using the indirect beam of ISN He. The upper left panel presents the correlation tube for the IBEX viewing conditions, repeated from Figure \ref{fig:vlCorrelTubesIBEX}, and a correlation tube obtained for the indirect beam of ISN He simulated for an interval of DOYs 305--321. The two tubes intersect at an angle sufficiently large for constraining the direction and speed of ISN He tightly. The upper right panel demonstrates that the magnitude of the indirect flux is within the capability of an IMAP-like instrument. Presented are the maxima of the indirect fluxes simulated for individual DOYSs for the three correlated parameter sets listed in the first three rows of Table \ref{tab:degenParam}. These fluxes are normalized to the maximum of the respective fluxes for the IBEX viewing  conditions. The lower row of panels presents normalized fluxes for the three correlated parameter sets, simulated for selected five DOYs from the indirect beam observation window. For these simulations, the normalization is done to the maximum of the fluxes for individual cases for the presented DOYs, i.e., it is different to that used in the upper-right panel. 
}
\label{fig:NdirCorrelTubes}
\end{figure}

For observations of ISN He performed close to the Earth's orbit, the indirect beam can be observed provided that the instrument boresight can be inclined to the Sun-centered spin axis of the spacecraft at least at 60\degr--75\degr. In principle, one could determine the inflow parameters of the ISN gas solely from observations of the indirect beam. However, the parameter correlation tubes are narrower for the indirect beam than for the direct beam observed by IBEX because of a relatively short observation arc. This is shown in the upper-left panel of Figure \ref{fig:NdirCorrelTubes}, where we repeat the IBEX correlation tube presented in Figure \ref{fig:vlCorrelTubesIBEX} and plot the correlation lines for two extreme DOYs when the whole indirect beam is visible for an instrument with the boresight elongation angle equal to 60\degr. Clearly, the correlation lines for the $\sim 15$ day observation window are almost identical. However, they strongly differ from the correlation lines for the easily-observable direct beam for DOYs only two months afterward. The intersection of the two correlation tubes will constrain the parameter magnitudes quite tightly.

The indirect beam was observed in the past by GAS/Ulysses \citep{witte:04}, but because of a  high physical background due to the Milky Way, helioglow, and stars it could not be used for precise determination of the inflow parameters. Here we argue that observing the indirect beam on a spacecraft like IMAP is feasible. 

First, the expected magnitude of the flux of the indirect beam is between 12\% and 20\% of the maximum flux for the IBEX viewing conditions even for the solar maximum epoch, as shown in the upper-right panel of Figure \ref{fig:NdirCorrelTubes}. The plot shows the expected flux normalized to the maximum of the respective fluxes for the IBEX viewing conditions. Such a flux magnitude is easily within the capability of an IBEX-like ISN instrument. Second, even though the correlation tubes obtained from the cold model are very close to each other, the actual beams with their thermal spread will not be challenging to differentiate, as shown in the lower row of panels of Figure \ref{fig:NdirCorrelTubes}. In these panels, we show normalized beams for the three correlated cases defined in the first three rows of Table \ref{tab:degenParam}. The shapes of the beams differ throughout the entire indirect beam window. We have verified that even though the Warm Breeze (i.e., the secondary population of ISN He) will be visible during the indicated time window, the WB signal is expected at a much lower level than that of the ISN beam, and consequently should not preclude using the indirect beam for the parameter correlation removal. The presented indirect beams for the tightly correlated parameters strongly differ from each other and thus are expected to be distinguishable in the actual observations despite the expected poorer statistics.

\subsection{Speed--temperature correlation}
\label{sec:speedTempCorrel}
\subsubsection{IBEX viewing conditions}
\label{sec:IBEXVTCorrel}
\noindent
Another correlation found in the IBEX data analysis and pointed out in Section \ref{sec:whyDegeneracy} is the correlation of the inflow speed and the gas temperature by the Mach number (Figure \ref{fig:TvCorrelTubesIBEX}). We simulated the expected IBEX-Lo signal for parameters correlated by the constant Mach number, as listed in the last two lines in Table \ref{tab:degenParam}. Here, the speed and the temperature are varied, but the inflow direction remains unchanged. These parameters are outside of the IBEX direction--speed correlation tube. 

\begin{figure}
\centering
\includegraphics[width=1.0\textwidth]{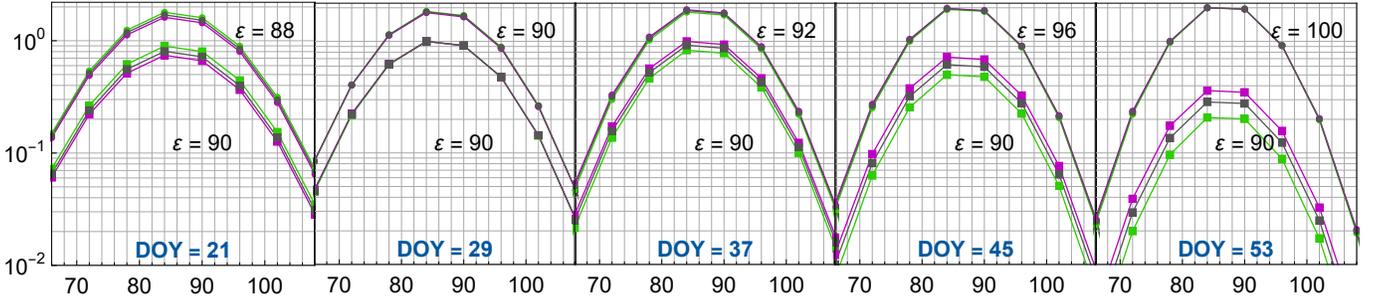}
\caption{
{\emph{ibexGeomTh}}
Two series of the normalized flux of ISN He for selected DOYs, corresponding to the IBEX ISN observation season, concerning the correlation between the speed and the temperature by the Mach number. The format is similar to that used in Figure \ref{fig:ibexGeom}. The figure illustrates the parameter correlation by the Mach number of the flow. The purple line corresponds to the parameters listed in the last (fifth) line in Table \ref{tab:degenParam}, the green line to those listed in the fourth line, and the black lines are identical to the black lines from Figure \ref{fig:ibexGeom}. }
\label{fig:ibexGeomTh}
\end{figure}

For the IBEX viewing conditions, i.e., with the elongation constant in time, the simulated signals are almost indistinguishable during a very short interval of time, but clearly different during the remaining portions of the IBEX observation season. This is illustrated in Figure \ref{fig:ibexGeomTh} in the lower set of lines. The difference in the signals for the three parameter sets are expected because the inflow parameters are outside of the experimentally-found parameter correlation tube. This suggests that it was possible to fit a unique set of the speed and temperature of the ISN gas based on IBEX observations owing to the fixed direction of the IBEX-Lo boresight direction. 

However, the speed--temperature correlation is real and does have consequences.  Simulations performed for the elongation angle varying to follow the peak of the ISN beam are hardly distinguishable, as demonstrated by the upper set of lines in Figure \ref{fig:ibexGeomTh} (which were scaled up by a factor of 2 to visually separate them from the other line set). In the following section, we will investigate how much this Mach number-related correlation persists in observations carried out along extended arcs around the Sun. 

\subsubsection{Removing the temperature--speed correlation}
\label{sec:IMAPVTCorrel}
\begin{figure}
\centering
\includegraphics[width=1.0\textwidth]{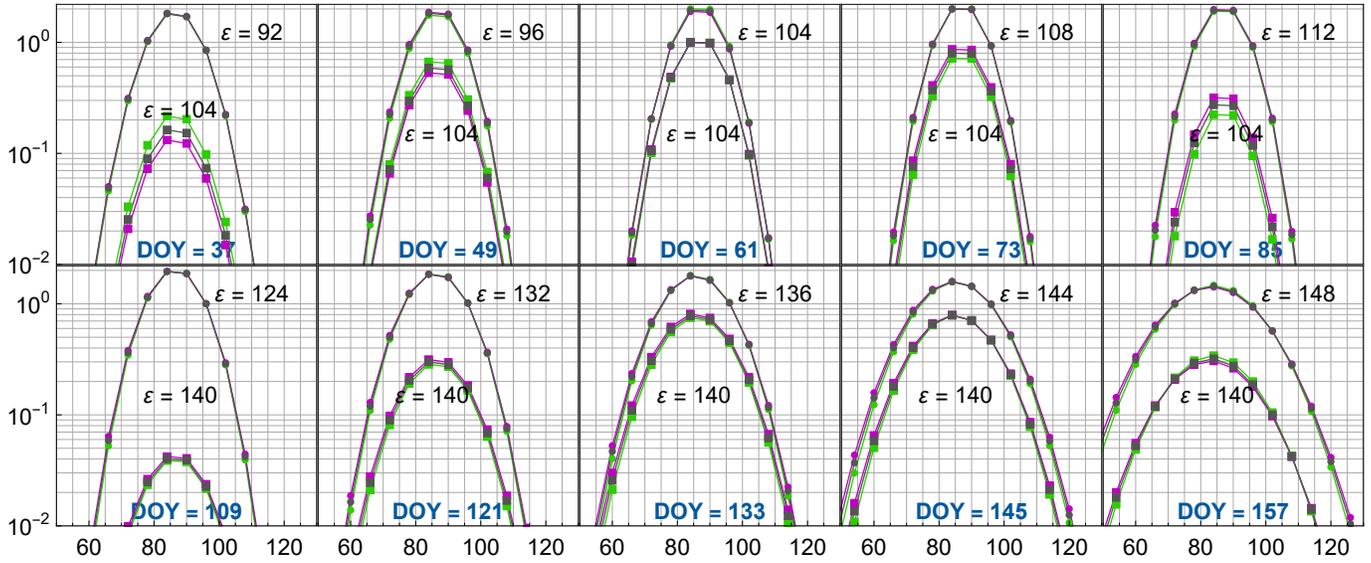}
\caption{
{\emph{followVsStepTh}}
Simulations of the normalized flux of ISN He for selected DOYs during the year for the three parameter sets correlated by the Mach number (lines 1, 4, and 5 in Table \ref{tab:degenParam}; black, green, and purple, respectively). The upper set of lines (dots) represents the ``follow the peak'' scenario (1) of adjusting the instrument boresight elongation angle $\varepsilon$, with the elongations listed in the individual panels. This set was scaled up by  a factor of 2 to visually separate it from the other set of simulations. Squares represent the ``stepwise adjustment'' scenario (2).}
\label{fig:followVsStepTh}
\end{figure}

\noindent
The Mach-number correlation was studied by simulations assuming the same scenarios for varying the boresight tilt angle as those used in Section \ref{sec:speedDirCorrel}. They are presented in Figure \ref{fig:followVsStepTh}. 

Clearly, removing the temperature-speed correlation within the ``follow the peak'' scenario (1) may be challenging, albeit feasible. The signals, represented by the upper set of figures (scaled by a factor of 2) are almost identical, and differences occur mostly away from the peaks. To remove the correlation within this scenario, it is recommended to use observations from time intervals distant by about three months (see the lines for DOY 157 and DOY 73) Note also that for a time around DOY 120, the direction--speed correlation coincides with the temperature-speed correlation (for the parameter sets listed in Table \ref{tab:degenParam}).

However, removing the temperature-speed correlation is expected to be much easier when using the stepwise adjustment scenario, as illustrated by the lower set of lines in Figure \ref{fig:followVsStepTh}. Even though  for the elongation angles used in this figure, the differences between the correlated cases seem smaller than those for the direction--speed correlated cases shown in Figure \ref{fig:followVsStep}, they can be increased by selecting a different elongation angle, as we have verified (simulations not shown). Nevertheless, the temperature--speed correlation is removed when one of these two parameters are determined independently, which can be done relatively easily with the speed, as discussed in Section \ref{sec:correlBreak}.

\subsection{Removing correlation of the inflow parameters for heavy species}
\label{sec:NeOxCorrel}
\noindent
Direct-sampling observations by IBEX-Lo revealed the presence of ISN O and Ne at the Earth's orbit \citep{mobius_etal:09b,bochsler_etal:12a}. While it is expected that Ne and O co-move with the ISN gas and that within the unperturbed LISM, the flow parameters of Ne and O are identical to those of ISN He, it cannot be ruled out that interactions within the outer heliosheath modify the populations of individual species in different ways \citep{schwadron_etal:16a,baliukin_etal:17a}. Therefore, it would be interesting to determine the flow parameters of the heavy interstellar species independently. 
\begin{figure}
\centering
\includegraphics[width=1.00\textwidth]{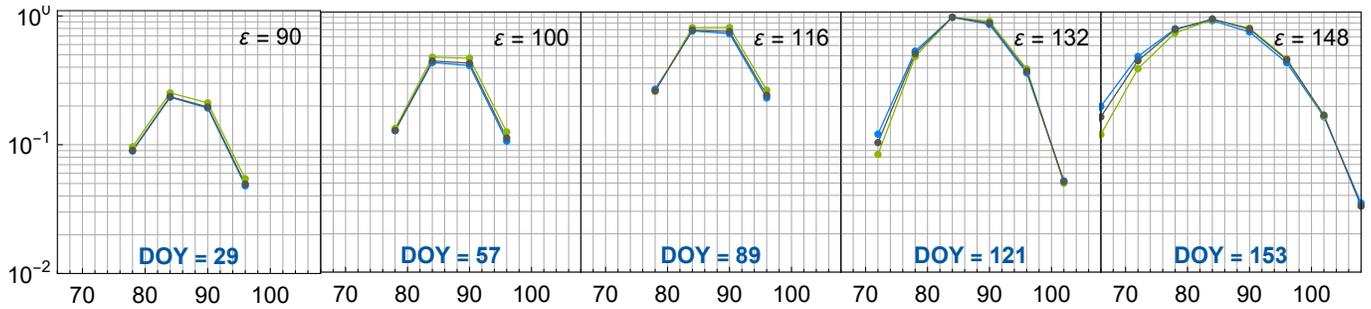}
\caption{
{\emph{followPeakOx}} 
Simulated flux of ISN O for selected DOYs and the correlated inflow parameters listed in the first three rows in Table \ref{tab:degenParam}. The format of the figure is identical to that of Figure \ref{fig:ibexGeom}. The fluxes for the three presented cases are normalized to their respective maximum values. A threshold of the absolute flux of 10 cm$^{-2}$ s$^{-1}$ was applied. The elongation angles were selected according to the ``follow the peak'' scenario.}
\label{fig:followPeakOx}
\end{figure}

Since Ne and O are practically insensitive to solar radiation pressure, they follow purely Keplerian trajectories inside the heliosphere, and the peaks of their beams at the Earth orbit are expected in the same locations in the spacecraft-inertial reference frame to those of ISN He. Since atomic masses of these species are much larger than that of He, the beam widths of Ne and O are much narrower than the beams of ISN He (compare Figure \ref{fig:followVsStep} and \ref{fig:followPeakOx}). And because  absolute densities of these species are estimated at $5.5 \times 10^{-6}$ \cc{} and $4.5\times 10^{-5}$ \cc, respectively \citep{frisch_slavin:03}, in contrast to that of ISN He at $1.5\times 10^{-3}$ \cc \citep{witte:04}, their fluxes at 1 au are expected to be lower by roughly three orders of magnitude than those of ISN He \citep{sokol_etal:19c}. This makes them a challenging target for direct-sampling observations. 

Despite these challenging conditions, Ne and O were detected in the IBEX-Lo data approximately at the expected signal level \citep{park_etal:14a,park_etal:15a,park_etal:19a}. It was found that the joint flux of Ne and O was lower by three orders of magnitude than that of ISN He, so the counting statistics actually obtained was much lower than that for this latter species. This, and technical issues in IBEX-Lo after 2011, prevented further observations of these species, but the data that had been collected enabled \citet{schwadron_etal:16a} to determine the flow parameters of ISN O. They were obtained similar to those found for ISN He, but with a much larger uncertainty. Also this latter study reported a strong correlation between the obtained flow parameters.

\begin{figure}
\centering
\includegraphics[width=1.00\textwidth]{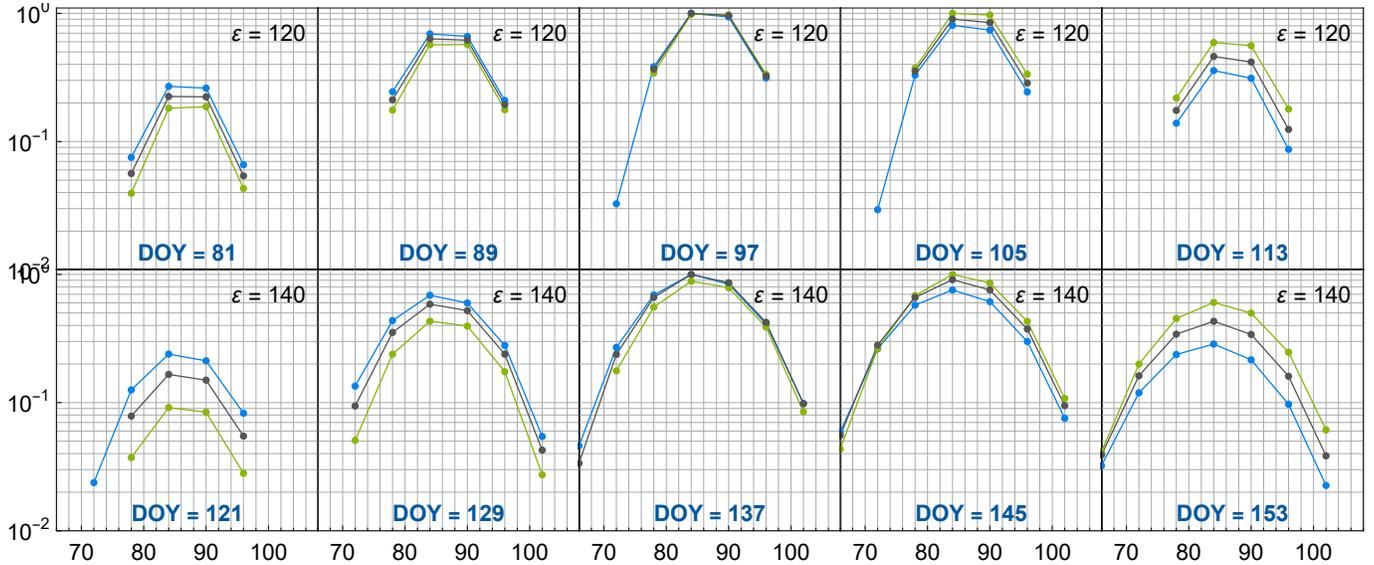}
\caption{
{\emph{stepOx}} 
Simulated flux of ISN O for selected DOYs and the correlated inflow parameters listed in the first three rows in Table \ref{tab:degenParam}. The format of the figure is identical to that of Figure \ref{fig:followVsStep}. The fluxes for the three presented cases are normalized to their respective maximum values. A threshold of the absolute flux of 10 cm$^{-2}$ s$^{-1}$ was applied. The elongation angles were selected according to a stepwise adjustment scenario with the elongation angles and DOYs listed at individual panels.}
\label{fig:stepOx}
\end{figure}

This is not surprising given the insight provided so far in this paper. The mechanism responsible for inflow parameter correlation is expected to be identical for all species. Therefore, it works also for Ne and O, even though it is challenging to separate O and Ne in the data because of the mechanism of neutral atom detection \citep[sputtering of O$^-$ ions by both O and Ne atoms,][]{wurz_etal:06a}.

We performed simulations for ISN O assuming the three alternative highly-correlated parameter sets listed in the first three rows in Table \ref{tab:degenParam}, assuming two alternative scenarios of adjusting the virtual instrument boresight: following the peak of the ISN beam, identical to that used in Figure \ref{fig:followVsStep}, and a stepwise modification, with a stepwise change of the boresight. Results of the first of them are presented in Figure \ref{fig:followPeakOx}. As in the preceding figures, we present the fluxes normalized to the maximum values of the presented time series, but to acknowledge the low absolute values of the expected flux, we introduced a low threshold of 10 atoms s$^{-1}$ cm$^{-2}$, approximately equal to the IBEX-Lo detection threshold \citep{park_etal:19a}.

The results illustrate that using the ``follow the peak'' scenario for boresight adjustment, one is able to collect measurements of ISN Ne and O with a low statistical scatter. Despite the small width of the beams, it will be possible to break the parameter correlation (note the different agreement levels for the simulations made with the correlated parameters for different DOYs, distributed along the first half of the year). The first of the presented panels approximately corresponds to the IBEX viewing conditions, the remaining four panels show the flux evolution and the increasing differences.

\begin{figure}
\centering
\includegraphics[width=0.5\textwidth]{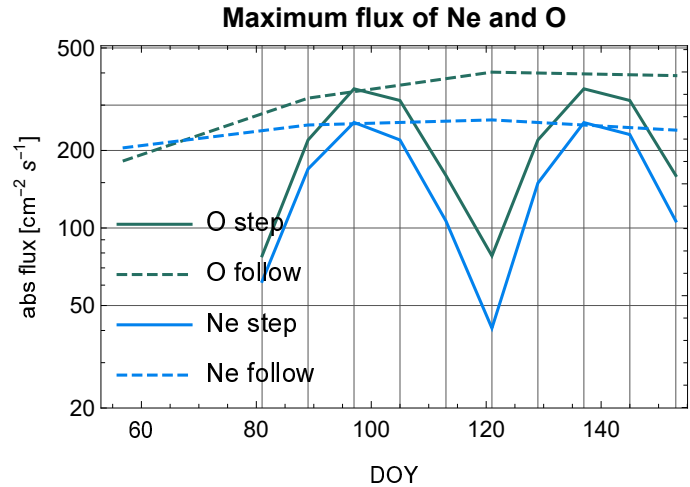}
\caption{
{\emph{fluxMaxPlot}} 
Maximum values of the absolute flux of the ISN Ne and O at an instrument orbiting the Sun at 1 au, simulated for selected DOYs assuming two alternative scenarios for adjusting the inclination of the instrument boresight to the Sun-oriented spacecraft spin axis: following the peak or stepwise, with the boresight elongation angle identical to that used in Figure \ref{fig:stepOx}.}
\label{fig:fluxMaxPlot} 
\end{figure}

Since, as discussed in Section \ref{sec:speedDirCorrel}, it is advantageous to use a stepwise boresight adjustment scenario for ISN He, we checked if adoption of such a scenario will be advantageous also in the case of ISN Ne and O. To that end, we performed simulations presented in Figure \ref{fig:stepOx}. The flux threshold is identical to that used in Figure \ref{fig:followPeakOx}.

The simulations clearly show that removal of parameter correlation for ISN Ne and O will be facilitated by adoption of the stepwise boresight adjustment scheme, similarly as in the case of ISN He. The differences between the correlated cases are larger than those expected when adopting the ``follow the peak'' scheme. 

Adopting the step-like boresight modification scheme is expected to provide provide measurements with a larger statistical scatter than that obtained using the ``follow the peak'' scheme. While this issue is not a significant drawback for the plentiful ISN He, it potentially might be an issue for Ne and O. However, analysis presented in Figure \ref{fig:fluxMaxPlot} demonstrates that adoption of a reasonable stepwise scheme, similar to that presented in Figure \ref{fig:stepOx}, can provide a sufficiently high signal to noise ratio to perform successful measurements and data analysis. The figure shows that the expected flux magnitudes are measurable, certainly not worse that those in the case of IBEX-Lo observations. The tested adjustment scheme offers two seasons of plentiful fluxes of ISN Ne and O, one about DOY 100, and another one about DOY 140, which is expected to provide two correlation tubes intersecting at large angles in the parameter space, as shown in Figure \ref{fig:CorrelTubesIMAP}. Thus, with the capabilities of IMAP-Lo, it will be possible to break the parameter correlation for the direction and speed of ISN Ne and O, not only for He. Breaking the speed--temperature correlation will be naturally achieved due to a tightly constrained speed magnitude. 

We note, however, that it is not likely that the option of using the indirect beams of the heavy species will be feasible, at least during the solar maximum conditions, because of a much larger attenuation of these beams by ionization than it is expected for helium. 

\subsection{Discussion}
\label{sec:discussion}
\noindent
The selection of the DOYs and elongation angles used in the example simulations presented in Section \ref{sec:thermalSpread} have not been optimized for the most advantageous elongation angles and fixed-elongation intervals. The ``stepwise adjustment'' scenario was adopted as an educated guess based on the insight from \citet{sokol_etal:19c}. However, it is evident from this study that breaking the parameter correlation is not difficult provided that observations are performed continuously or intermittently, but during time intervals distant by a couple of months during the year. If the adopted elongation angles allow a sufficiently large flux of ISN He into the instrument, the correlation will be removed, and a scheme of elongation adjustment can be defined to accommodate also other science targets, as those discussed in \citet{sokol_etal:19c}.

The discussion presented so far focused on the primary population of the ISN gas. However, identical parameter correlation mechanisms operate also for the secondary population, which, similarly to the primary population, also are obtained strongly correlated \citep{kubiak_etal:16a}. The proposed ideas to break the parameter correlation are applicable in studies of the secondary population as well.

\section{Summary and conclusions}
\label{sec:conclu}
\noindent
We have investigated the reasons for the correlation of the inflow parameters of the ISN gas, obtained in analyzes of direct-sampling observations performed by the IBEX-Lo experiment. The presence of the correlation is visible as ``tubes'' of the more likely values of the parameters in the parameter space. 

We found that the parameter correlation appears because of physical reasons. As long as there is no capability of measuring both the direction and the speed of the atoms incoming at the instrument, and only the direction can be measured, the flow direction, the speed, and the temperature of the ISN gas will be obtained correlated, when observations are carried out along a short orbital arc of the instrument. An illustration of the reason is presented in Figure \ref{fig:degenOrbits}.

This conclusion is obtained from a simple reasoning, presented in Section \ref{sec:whyDegeneracy}, supported by calculations performed using the cold model of the ISN gas flow. With this model, we approximately reproduced the parameter tube obtained from IBEX observations, as shown in Figure \ref{fig:vlCorrelTubesIBEX}. This conclusion is also supported experimentally, by analysis of the ISN He flow parameters based on measurements of the Ulysses/GAS by \citet{bzowski_etal:14a}, discussed in Section \ref{sec:whyDegeneracy}. 

The parameter correlation will not appear in the analysis if the observations are performed along a sufficiently long orbital arc around the Sun, or at least over shorter orbital arcs of the instrument, separated in space. The orientation of a parameter tube in the parameter space varies with the location along the orbit. When the instrument moves with the spacecraft along the orbit, the parameter tube rotates in the parameter space, and the daily tubes intersect at the position of the actual inflow parameters. This is illustrated in Figure \ref{fig:CorrelTubesIMAP}. By using this feature, it is easy to break the correlation between the direction and the speed of the inflow of ISN gas.

Another option to break the parameter correlation, complementary to the aforementioned one, is using observations of both the direct and indirect beams of ISN He, as discussed in Section \ref{sec:ndirBeam}. This requires using elongation angles of the instrument boresight about 60\degr--75\degr{} and is feasible only for He, because of large ionization losses for O and Ne. 

The correlation between the speed and the temperature appears because combinations of the speed and the temperature corresponding to identical Mach numbers of the flow result in identical widths of the ISN gas beam. The speed--temperature tube is shown in Figure \ref{fig:TvCorrelTubesIBEX}. In the simple terms of the cold model, this correlation persists along the orbit around the Sun.

By simulations performed using the hot model of the ISN gas flow in the heliosphere, implemented in the WTPM model, we showed that indeed, the simulated beams of the ISN gas observed by an IBEX-like instrument, performed for alternative parameter sets correlated by speed and direction for a given location in the orbit, predicted by the cold model, are almost indistinguishable, as shown in Figure \ref{fig:ibexGeom}. We showed that the same parameter sets are outside the parameter tube for different location along the Earth orbit, which results in strongly differing simulated beams, presented in Figure \ref{fig:followVsStep}. Such beams would be easy to differentiate in actual observations. This supports the idea that performing the observations along sufficiently long orbital arcs enables breaking the parameter correlation.

Breaking the speed-temperature correlation will also be possible, since the similar beams obtained for the correlated parameters in one location will become different for a different location, as shown in Figure \ref{fig:followVsStepTh}. 

The ability to observe the ISN gas from different locations along the Earth orbit requires a capability to adjust the boresight of the instrument, i.e., the elongation angle between the boresight and the spacecraft spin axis, such as that of the planned IMAP-Lo experiment. In Figures \ref{fig:followVsStep} and \ref{fig:followVsStepTh} we show that successful parameter breaking is achievable for different scenarios of adjusting the elongation angle as the measurement location moves along the Earth orbit. This is true not only for the case when the boresight is adjusted to follow the peak of the ISN gas. In fact, adopting a scenario with the boresight elongation maintained constant during prolonged intervals may be more advantageous.

We verified this conclusion not only for the planned observations of the plentiful ISN He, but also for those of ISN Ne and O, which are less abundant and hence more challenging to observe and interpret. In section \ref{sec:NeOxCorrel} we show simulations for these species and point out that the expected statistics collected by an IMAP-like instrument is expected to be sufficient when adopting both the peak-following and the stepwise boresight adjustment scenarios. 

Summarizing: the parameter correlation is due to physical reasons and mere increasing the statistics while maintaining the location of observations in the Earth's orbit will not remove it. Breaking the ISN gas parameter correlation requires performing observations from separate locations in the orbit relatively far apart, and satisfying this prerequisite requires a capability of adjusting the instrument boresight relative to the spacecraft rotation axis, such as that of the planned IMAP-Lo camera.

\begin{acknowledgments}
 The work at CBK PAN was supported by the Polish National Science Centre grant 2019/35/B/ST9/01241.
\end{acknowledgments}
\bibliographystyle{aasjournal}
\bibliography{breakDege_v5}

\begin{thebibliography}{}
\expandafter\ifx\csname natexlab\endcsname\relax\def\natexlab#1{#1}\fi
\providecommand{\url}[1]{\href{#1}{#1}}
\providecommand{\dodoi}[1]{doi:~\href{http://doi.org/#1}{\nolinkurl{#1}}}
\providecommand{\doeprint}[1]{\href{http://ascl.net/#1}{\nolinkurl{http://ascl.net/#1}}}
\providecommand{\doarXiv}[1]{\href{https://arxiv.org/abs/#1}{\nolinkurl{https://arxiv.org/abs/#1}}}

\bibitem[{Axford {et~al.}(1963)Axford, Dessler, \& Gottlieb}]{axford_etal:63a}
Axford, W., Dessler, A., \& Gottlieb, B. 1963, \apj, 137, 1268

\bibitem[{Axford(1972)}]{axford:72}
Axford, W.~I. 1972, in The Solar Wind, ed. J.~M.~W. C.~P.~Sonnet, P.
  J.~Coleman, NASA Spec. Publ. 308, 609--660

\bibitem[{{Baliukin} {et~al.}(2017){Baliukin}, {Izmodenov}, {M{\"o}bius},
  {Alexashov}, {Katushkina}, \& {Kucharek}}]{baliukin_etal:17a}
{Baliukin}, I.~I., {Izmodenov}, V.~V., {M{\"o}bius}, E., {et~al.} 2017, \apj,
  850, 119, \dodoi{10.3847/1538-4357/aa93e8}

\bibitem[{Baranov {et~al.}(1991)Baranov, Lebedev, \& Malama}]{baranov_etal:91}
Baranov, V.~B., Lebedev, M.~G., \& Malama, Y.~G. 1991, \apj, 375, 347

\bibitem[{Bertaux \& Blamont(1971)}]{bertaux_blamont:71}
Bertaux, J.~L., \& Blamont, J.~E. 1971, \aap, 11, 200

\bibitem[{Blum \& Fahr(1970)}]{blum_fahr:70a}
Blum, P., \& Fahr, H.~J. 1970, \aap, 4, 280

\bibitem[{Bochsler {et~al.}(2012)Bochsler, Petersen, M{\"o}bius, Kucharek,
  Schwadron, Wurz, Saul, Scheer, Fuselier, McComas, Bzowski, \&
  Frisch}]{bochsler_etal:12a}
Bochsler, P., Petersen, L., M{\"o}bius, E., {et~al.} 2012, \apjs, 198, 13,
  \dodoi{10.1088/0067-0049/198/2/13}

\bibitem[{{Bzowski} {et~al.}(2017){Bzowski}, {Kubiak}, {Czechowski}, \&
  {Grygorczuk}}]{bzowski_etal:17a}
{Bzowski}, M., {Kubiak}, M.~A., {Czechowski}, A., \& {Grygorczuk}, J. 2017,
  \apj, 845, 15.
\newblock \doarXiv{1707.02193}

\bibitem[{{Bzowski} {et~al.}(2014){Bzowski}, {Kubiak}, {H{\l}ond},
  {Sok{\'o}{\l}}, {Banaszkiewicz}, \& {Witte}}]{bzowski_etal:14a}
{Bzowski}, M., {Kubiak}, M.~A., {H{\l}ond}, M., {et~al.} 2014, \aap, 569, A8,
  \dodoi{10.1051/0004-6361/201424127}

\bibitem[{Bzowski {et~al.}(2013)Bzowski, Sok{\'o}{\l}, Kubiak, \&
  Kucharek}]{bzowski_etal:13b}
Bzowski, M., Sok{\'o}{\l}, J.~M., Kubiak, M.~A., \& Kucharek, H. 2013, \aap,
  557, A50, \dodoi{10.1051/0004-6361/201321700}

\bibitem[{Bzowski {et~al.}(2012)Bzowski, Kubiak, M{\"o}bius, Bochsler, Leonard,
  Heirtzler, Kucharek, Sok{\'{o}}{\l}, H{\l}ond, Crew, Schwadron, Fuselier, \&
  McComas}]{bzowski_etal:12a}
Bzowski, M., Kubiak, M.~A., M{\"o}bius, E., {et~al.} 2012, \apjs, 198, 12,
  \dodoi{10.1088/0067-0049/198/2/12}

\bibitem[{{Bzowski} {et~al.}(2015){Bzowski}, {Swaczyna}, {Kubiak},
  {Sok\'{o}{\l}}, {Fuselier}, {Galli}, {Heirtzler}, {Kucharek}, {Leonard},
  {McComas}, {M{\"o}bius}, {Schwadron}, \& {Wurz}}]{bzowski_etal:15a}
{Bzowski}, M., {Swaczyna}, P., {Kubiak}, M.~A., {et~al.} 2015, \apjs, 220, 28,
  \dodoi{10.1088/0067-0049/220/2/28}

\bibitem[{Bzowski {et~al.}(2019)Bzowski, Czechowski, Frisch, Fuselier, Galli,
  Grygorczuk, Heerikhuisen, Kubiak, Kucharek, McComas, M{\"o}bius, Schwadron,
  Slavin, Sok{\'o}{\l}, Swaczyna, Wurz, \& Zirnstein}]{bzowski_etal:19a}
Bzowski, M., Czechowski, A., Frisch, P., {et~al.} 2019, \apj, 882, 60,
  \dodoi{10.3847/1538-4357/ab3462}

\bibitem[{Fahr(1968)}]{fahr:68}
Fahr, H.~J. 1968, \apss, 2, 474

\bibitem[{{Fahr}(1974)}]{fahr:74}
{Fahr}, H.~J. 1974, \ssr, 15, 483, \dodoi{10.1007/BF00178217}

\bibitem[{{Fahr} \& {Lay}(1973)}]{fahr_lay:73a}
{Fahr}, H.~J., \& {Lay}, G. 1973, in Space Research Conference, Vol.~2, Space
  Research Conference, ed. M.~J. {Rycroft} \& S.~K. {Runcorn}, 843--847

\bibitem[{Fraternale {et~al.}(2021)Fraternale, Pogorelov, \&
  Heerikhuisen}]{fraternale_etal:21a}
Fraternale, F., Pogorelov, N.~V., \& Heerikhuisen, J. 2021, \apjl, 921, L24,
  \dodoi{10.3847/2041-8213/ac313c}

\bibitem[{{Frisch} \& {Slavin}(2003)}]{frisch_slavin:03}
{Frisch}, P.~C., \& {Slavin}, J.~D. 2003, \apj, 594, 844

\bibitem[{{Fuselier} {et~al.}(2009){Fuselier}, {Bochsler}, {Chornay}, {Clark},
  {Crew}, {Dunn}, {Ellis}, {Friedmann}, {Funsten}, {Ghielmetti}, {Googins},
  {Granoff}, {Hamilton}, {Hanley}, {Heirtzler}, {Hertzberg}, {Isaac}, {King},
  {Knauss}, {Kucharek}, {Kudirka}, {Livi}, {Lobell}, {Longworth}, {Mashburn},
  {McComas}, {M{\"o}bius}, {Moore}, {Moore}, {Nemanich}, {Nolin}, {O'Neal},
  {Piazza}, {Peterson}, {Pope}, {Rosmarynowski}, {Saul}, {Scherrer}, {Scheer},
  {Schlemm}, {Schwadron}, {Tillier}, {Turco}, {Tyler}, {Vosbury}, {Wieser},
  {Wurz}, \& {Zaffke}}]{fuselier_etal:09b}
{Fuselier}, S.~A., {Bochsler}, P., {Chornay}, D., {et~al.} 2009, \ssr, 146,
  117, \dodoi{10.1007/s11214-009-9495-8}

\bibitem[{{Galli} {et~al.}(2015){Galli}, {Wurz}, {Park}, {Kucharek},
  {M{\"o}bius}, {Schwadron}, {Sok\'{o}{\l}}, {Bzowski}, {Kubiak}, {Swaczyna},
  {Fuselier}, \& {McComas}}]{galli_etal:15a}
{Galli}, A., {Wurz}, P., {Park}, J., {et~al.} 2015, \apjs, 220, 30,
  \dodoi{10.1088/0067-0049/220/2/30}

\bibitem[{{Galli} {et~al.}(2019){Galli}, {Wurz}, {Rahmanifard}, {M{\"o}bius},
  {Schwadron}, {Kucharek}, {Heirtzler}, {Fairchild}, {Bzowski}, {Kubiak},
  {Kowalska-Leszczy{\'n}ska}, {Sok{\'o}{\l}}, {Fuselier}, {Swaczyna}, \&
  {McComas}}]{galli_etal:19a}
{Galli}, A., {Wurz}, P., {Rahmanifard}, F., {et~al.} 2019, \apj, 871, 52,
  \dodoi{10.3847/1538-4357/aaf737}

\bibitem[{Holzer(1977)}]{holzer:77}
Holzer, T.~E. 1977, Rev. Geophys., 15, 467

\bibitem[{{Johnson}(1972{\natexlab{a}})}]{johnson:72a}
{Johnson}, H.~E. 1972{\natexlab{a}}, \planss, 20, 829,
  \dodoi{10.1016/0032-0633(72)90168-7}

\bibitem[{{Johnson}(1972{\natexlab{b}})}]{johnson:72b}
---. 1972{\natexlab{b}}, \planss, 20, 1784,
  \dodoi{10.1016/0032-0633(72)90199-7}

\bibitem[{Katushkina {et~al.}(2015)Katushkina, Izmodenov, Alexashov, Schwadron,
  \& McComas}]{katushkina_etal:15b}
Katushkina, O.~A., Izmodenov, V.~V., Alexashov, D.~B., Schwadron, N.~A., \&
  McComas, D.~J. 2015, \apjs, 220, 33, \dodoi{10.1088/0067-0049-220-2-33}

\bibitem[{{Kubiak} {et~al.}(2014){Kubiak}, {Bzowski}, {Sok{\'o}{\l}},
  {Swaczyna}, {Grzedzielski}, {Alexashov}, {Izmodenov}, {Moebius}, {Leonard},
  {Fuselier}, {Wurz}, \& {McComas}}]{kubiak_etal:14a}
{Kubiak}, M.~A., {Bzowski}, M., {Sok{\'o}{\l}}, J.~M., {et~al.} 2014, \apjs,
  213, 29, \dodoi{10.1088/0067-0049/212/2/29}

\bibitem[{Kubiak {et~al.}(2016)Kubiak, Swaczyna, Bzowski, Sok{\'o}{\l},
  Fuselier, Galli, Heirtzler, Kucharek, Leonard, McComas, Park, Schwadron, \&
  Wurz}]{kubiak_etal:16a}
Kubiak, M.~A., Swaczyna, P., Bzowski, M., {et~al.} 2016, \apjs, 223, 35,
  \dodoi{10.1088/0067-0049/220/2/35}

\bibitem[{Lee {et~al.}(2015)Lee, M{\"o}bius, \& Leonard}]{lee_etal:15a}
Lee, M.~A., M{\"o}bius, E., \& Leonard, T.~W. 2015, \apjs, 220, 23,
  \dodoi{10.1088/0067-0049/220/2/23}

\bibitem[{McComas {et~al.}(2018)McComas, Christian, Schwadron, Fox, Westlake,
  Allegrini, Baker, Biesecker, Bzowski, Clark, Cohen, Cohen, Dayeh, Decker,
  de~Nolfo, Desai, andH.A. Elliott, Fahr, Frisch, Funsten, Fuselier, Galli,
  Galvin, Giacalone, Gkioulidou, Guo, Horanyi, Isenberg, Janzen, Kistler,
  Korreck, Kubiak, Kucharek, Larsen, Leske, Lugaz, Luhmann, Matthaeus, Mitchel,
  Moebius, Ogasawara, Reisenfeld, Richardson, Russell, Sok{\'o}{\l}, Spence,
  Skoug, Sternovsky, Swaczyna, Szalay, Tokumaru, andP. Wurz, Zank, \&
  Zirnstein}]{mccomas_etal:18b}
McComas, D., Christian, E., Schwadron, N., {et~al.} 2018, \ssr, 214, 116,
  \dodoi{10.1007/s11214-018-0550-1}

\bibitem[{{McComas} {et~al.}(2009){McComas}, {Allegrini}, {Bochsler},
  {Bzowski}, {Collier}, {Fahr}, {Fichtner}, {Frisch}, {Funsten}, {Fuselier},
  {Gloeckler}, {Gruntman}, {Izmodenov}, {Knappenberger}, {Lee}, {Livi},
  {Mitchell}, {M{\"o}bius}, {Moore}, {Pope}, {Reisenfeld}, {Roelof},
  {Scherrer}, {Schwadron}, {Tyler}, {Wieser}, {Witte}, {Wurz}, \&
  {Zank}}]{mccomas_etal:09a}
{McComas}, D.~J., {Allegrini}, F., {Bochsler}, P., {et~al.} 2009, \ssr, 146,
  11, \dodoi{10.1007/s11214-009-9499-4}

\bibitem[{{McComas} {et~al.}(2011){McComas}, {Carrico}, {Hautamaki},
  {Intelisano}, {Lebois}, {Loucks}, {Policastri}, {Reno}, {Scherrer},
  {Schwadron}, {Tapley}, \& {Tyler}}]{mccomas_etal:11a}
{McComas}, D.~J., {Carrico}, J.~P., {Hautamaki}, B., {et~al.} 2011, Space
  Weather, 9, S11002, \dodoi{10.1029/2011SW000704}

\bibitem[{{M{\"o}bius} {et~al.}(2015{\natexlab{a}}){M{\"o}bius}, {Lee}, \&
  {Drews}}]{mobius_etal:15c}
{M{\"o}bius}, E., {Lee}, M.~A., \& {Drews}, C. 2015{\natexlab{a}}, \apj, 815,
  \dodoi{10.1088/0004-637X/815/1/20}

\bibitem[{{M{\"o}bius} {et~al.}(2009{\natexlab{a}}){M{\"o}bius}, {Kucharek},
  {Clark}, {O'Neill}, {Petersen}, {Bzowski}, {Saul}, {Wurz}, {Fuselier},
  {Izmodenov}, {McComas}, {M{\"u}ller}, \& {Alexashov}}]{mobius_etal:09a}
{M{\"o}bius}, E., {Kucharek}, H., {Clark}, G., {et~al.} 2009{\natexlab{a}},
  \ssr, 146, 149, \dodoi{10.1007/s11214-009-9498-5}

\bibitem[{{M{\"o}bius} {et~al.}(2009{\natexlab{b}}){M{\"o}bius}, {Bochsler},
  {Bzowski}, {Crew}, {Funsten}, {Fuselier}, {Ghielmetti}, {Heirtzler},
  {Izmodenov}, {Kubiak}, {Kucharek}, {Lee}, {Leonard}, {McComas}, {Petersen},
  {Saul}, {Scheer}, {Schwadron}, {Witte}, \& {Wurz}}]{mobius_etal:09b}
{M{\"o}bius}, E., {Bochsler}, P., {Bzowski}, M., {et~al.} 2009{\natexlab{b}},
  Science, 326, 969, \dodoi{10.1126/science.1180971}

\bibitem[{M{\"o}bius {et~al.}(2012)M{\"o}bius, Bochsler, Heirtzler, Kucharek,
  Lee, Leonard, Petersen, Schwadron, Valocvin, Wu, Bzowski, Kubiak, Fuselier,
  Saul, Wurz, McComas, \& Crew}]{mobius_etal:12a}
M{\"o}bius, E., Bochsler, P., Heirtzler, D., {et~al.} 2012, \apjs, 198, 11,
  \dodoi{10.1088/0067-0049/198/2/11}

\bibitem[{{M{\"o}bius} {et~al.}(2015{\natexlab{b}}){M{\"o}bius}, {Bzowski},
  {Fuselier}, {Heirtzler}, {Kubiak}, {Kucharek}, {Lee}, {Leonard}, {McComas},
  {Schwadron}, {Sok\'{o}{\l}}, \& {Wurz}}]{mobius_etal:15b}
{M{\"o}bius}, E., {Bzowski}, M., {Fuselier}, S.~A., {et~al.}
  2015{\natexlab{b}}, \apjs, 220, 24, \dodoi{10.1088/0067-0049/220/2/24}

\bibitem[{{Morton} \& {Purcell}(1962)}]{morton_purcell:62a}
{Morton}, D.~C., \& {Purcell}, J.~D. 1962, \planss, 9, 455,
  \dodoi{10.1016/0032-0633(62)90048-X}

\bibitem[{Park {et~al.}(2015)Park, Kucharek, M{\"o}bius, Galli, Livadiotis,
  Fuselier, \& J.McComas}]{park_etal:15a}
Park, J., Kucharek, H., M{\"o}bius, E., {et~al.} 2015, \apjs, 220,
  \dodoi{10.1088/0067-0049-220-2-34}

\bibitem[{Park {et~al.}(2019)Park, Kucharek, Pachalidis, Szabo, Heirtzler, \&
  M{\"o}bius}]{park_etal:19a}
Park, J., Kucharek, H., Pachalidis, N., {et~al.} 2019, \apj, 880, 4,
  \dodoi{10.3847/1538-4357/ab264a}

\bibitem[{Park {et~al.}(2014)Park, Kucharek, M{\"o}bius, Leonard, Bzowski,
  Sok{\'o}{\l}, Kubiak, Fuselier, \& J.McComas}]{park_etal:14a}
Park, J., Kucharek, H., M{\"o}bius, E., {et~al.} 2014, \apj, 795, 97,
  \dodoi{10.1088/0004-637X/795/1/97}

\bibitem[{Patterson {et~al.}(1963)Patterson, Johnson, \&
  Hanson}]{patterson_etal:63a}
Patterson, T., Johnson, F., \& Hanson, W. 1963, \planss, 11, 767,
  \dodoi{10.1016/0032-0633(63)90189-2}

\bibitem[{Rahmanifard {et~al.}(2019)Rahmanifard, M{\"o}bius, Schwadron, Galli,
  Richards, Kucharek, Sok{\'{o}}{\l}, Heirtzler, Lee, Bzowski,
  Kowalska-Leszczynska, Kubiak, Wurz, Fuselier, \&
  McComas}]{rahmanifard_etal:19a}
Rahmanifard, F., M{\"o}bius, E., Schwadron, N.~A., {et~al.} 2019, \apj, 887,
  217, \dodoi{10.3847/1538-4357/ab58ce}

\bibitem[{{Rodr{\'i}guez Moreno} {et~al.}(2013){Rodr{\'i}guez Moreno}, Wurz,
  Saul, Bzowski, Kubiak, Sok{\'o\l}, Frisch, Fuselier, McComas, M{\"o}bius, \&
  Schwadron}]{rodriguez_etal:13a}
{Rodr{\'i}guez Moreno}, D.~F., Wurz, P., Saul, L., {et~al.} 2013, \aap, 557,
  A125, \dodoi{10.1051/0004-6361/201321420}

\bibitem[{Ruci{\'n}ski {et~al.}(1996)Ruci{\'n}ski, Cummings, Gloeckler,
  Lazarus, M{\"o}bius, \& Witte}]{rucinski_etal:96a}
Ruci{\'n}ski, D., Cummings, A.~C., Gloeckler, G., {et~al.} 1996, \ssr, 78, 73,
  \dodoi{10.1007/BF00170794}

\bibitem[{Saul {et~al.}(2012)Saul, Wurz, M{\"o}bius, Bzowski, Fuselier, Crew,
  Rodriguez, Leonard, McComas, Schwadron, Bochsler, \& Scheer}]{saul_etal:12a}
Saul, L., Wurz, P., M{\"o}bius, E., {et~al.} 2012, \apjs, 198, 14,
  \dodoi{10.1088/0067-0049/198/2/14}

\bibitem[{{Schwadron} {et~al.}(2015){Schwadron}, {M{\"o}bius}, {Leonard},
  {Fuselier}, {Bzowski}, {Frisch}, {Heirtzler}, {Kubiak}, {Kucharek}, {Lee},
  {McComas}, {Rahmanifard}, {Sok\'{o}{\l}}, \& {Swaczyna}}]{schwadron_etal:15a}
{Schwadron}, N., {M{\"o}bius}, E., {Leonard}, T., {et~al.} 2015, \apjs, 220,
  25, \dodoi{10.1088/0067-0049/220/2/25}

\bibitem[{{Schwadron} {et~al.}(2013){Schwadron}, {Moebius}, {Kucharek}, {Lee},
  {French}, {Saul}, {Wurz}, {Bzowski}, {Fuselier}, {Livadiotis}, {McComas},
  {Frisch}, {Gruntman}, \& {Mueller}}]{schwadron_etal:13a}
{Schwadron}, N.~A., {Moebius}, E., {Kucharek}, H., {et~al.} 2013, \apj, 775,
  86, \dodoi{10.1088/0004-637X/775/2/86}

\bibitem[{{Schwadron} {et~al.}(2016){Schwadron}, {M{\"o}bius}, {McComas},
  {Bochsler}, {Bzowski}, {Fuselier}, {Livadiotis}, {Frisch}, {M{\"u}ller},
  {Heirtzler}, {Kucharek}, \& {Lee}}]{schwadron_etal:16a}
{Schwadron}, N.~A., {M{\"o}bius}, E., {McComas}, D.~J., {et~al.} 2016, \apj,
  828, 81, \dodoi{10.3847/0004-637X/828/2/81}

\bibitem[{Schwadron {et~al.}(2022)Schwadron, M{\"o}bius, McComas, Bower, Bower,
  Bzowski, Fuselier, Heirtzler, Kubiak, Lee, Rahmanifard, Sok{\'o}{\l},
  Swaczyn, \& Winslow}]{schwadron_etal:22a}
Schwadron, N.~A., M{\"o}bius, E., McComas, D.~J., {et~al.} 2022, \apjs, 258, 7,
  \dodoi{10.3847/1538-4365/ac2fa9}

\bibitem[{{Slavin} \& {Frisch}(2007)}]{slavin_frisch:07a}
{Slavin}, J.~D., \& {Frisch}, P.~C. 2007, \ssr, 130, 409,
  \dodoi{10.1007/s11214-007-9186-2}

\bibitem[{{Sok{\'o}{\l}} {et~al.}(2019{\natexlab{a}}){Sok{\'o}{\l}}, {Bzowski},
  \& {Tokumaru}}]{sokol_etal:19a}
{Sok{\'o}{\l}}, J.~M., {Bzowski}, M., \& {Tokumaru}, M. 2019{\natexlab{a}},
  \apj, 872, 57, \dodoi{10.3847/1538-4357/aaf737}

\bibitem[{{Sok{\'o}{\l}} {et~al.}(2019{\natexlab{b}}){Sok{\'o}{\l}}, {Kubiak},
  {Bzowski}, {M{\"o}bius}, \& {Schwadron}}]{sokol_etal:19c}
{Sok{\'o}{\l}}, J.~M., {Kubiak}, M.~A., {Bzowski}, M., {M{\"o}bius}, E., \&
  {Schwadron}, N. 2019{\natexlab{b}}, \apjs, 245, 28,
  \dodoi{10.3847/1538-4365/ab50bc}

\bibitem[{{Sok\'{o}{\l}} {et~al.}(2015{\natexlab{a}}){Sok\'{o}{\l}}, {Kubiak},
  {Bzowski}, \& {Swaczyna}}]{sokol_etal:15b}
{Sok\'{o}{\l}}, J.~M., {Kubiak}, M.~A., {Bzowski}, M., \& {Swaczyna}, P.
  2015{\natexlab{a}}, \apjs, 220, 27, \dodoi{10.1088/0067-0049/220/2/27}

\bibitem[{{Sok{\'o}{\l}} {et~al.}(2020){Sok{\'o}{\l}}, {McComas}, {Bzowski}, \&
  {Tokumaru}}]{sokol_etal:20a}
{Sok{\'o}{\l}}, J.~M., {McComas}, D.~J., {Bzowski}, M., \& {Tokumaru}, M. 2020,
  \apj, 897, 179, \dodoi{10.3847/1538-4357/ab99a4}

\bibitem[{{Sok\'{o}{\l}} {et~al.}(2015{\natexlab{b}}){Sok\'{o}{\l}}, {Bzowski},
  {Kubiak}, {Swaczyna}, {Galli}, {Wurz}, {M{\"o}bius}, {Kucharek}, {Fuselier},
  \& {McComas}}]{sokol_etal:15a}
{Sok\'{o}{\l}}, J.~M., {Bzowski}, M., {Kubiak}, M.~A., {et~al.}
  2015{\natexlab{b}}, \apjs, 220, 29, \dodoi{10.1088/0067-0049/220/2/29}

\bibitem[{Swaczyna {et~al.}(2021)Swaczyna, Rahmanifard, Zirnstein, McComas, \&
  Heerikhuisen}]{swaczyna_etal:21a}
Swaczyna, P., Rahmanifard, F., Zirnstein, E., McComas, D., \& Heerikhuisen, J.
  2021, \apjl, 911, L48, \dodoi{10.3847/2041-8213/abf436}

\bibitem[{{Swaczyna} {et~al.}(2015){Swaczyna}, {Bzowski}, {Kubiak},
  {Sok\'{o}{\l}}, {M{\"o}bius}, {Leonard}, {Heirtzler}, {Kucharek},
  {Schwadron}, {Fuselier}, \& {McComas}}]{swaczyna_etal:15a}
{Swaczyna}, P., {Bzowski}, M., {Kubiak}, M.~A., {et~al.} 2015, \apjs, 220, 26,
  \dodoi{10.1088/0067-0049/220/2/26}

\bibitem[{Swaczyna {et~al.}(2018)Swaczyna, Bzowski, Kubiak, Sok{\'o}{\l},
  Fuselier, Galli, Heirtzler, Kucharek, McComas, M{\"o}bius, Schwadron, \&
  Wurz}]{swaczyna_etal:18a}
Swaczyna, P., Bzowski, M., Kubiak, M.~A., {et~al.} 2018, \apj, 854, 119,
  \dodoi{10.3847/1538-4357/aaabbf}

\bibitem[{Swaczyna {et~al.}(2022)Swaczyna, Kubiak, Bzowski, Bower, Fuselier,
  Galli, Heirtzler, McComas, M{\"o}bius, Rahmanifard, \&
  Schwadron}]{swaczyna_etal:22b}
Swaczyna, P., Kubiak, M.~A., Bzowski, M., {et~al.} 2022, \apjs, 259, 42,
  \dodoi{10.3847/1538-4365/ac4bde}

\bibitem[{{Taut} {et~al.}(2018){Taut}, {Berger}, {M{\"o}bius}, {Drews},
  {Heidrich-Meisner}, {Keilbach}, {Lee}, \&
  {Wimmer-Schweingruber}}]{taut_etal:18a}
{Taut}, A., {Berger}, L., {M{\"o}bius}, E., {et~al.} 2018, \aap, 611, A61,
  \dodoi{10.1051/0004-6361/201731796}

\bibitem[{Thomas(1978)}]{thomas:78}
Thomas, G.~E. 1978, Ann. Rev. Earth Planet. Sci., 6, 173

\bibitem[{Thomas \& Krassa(1971)}]{thomas_krassa:71}
Thomas, G.~E., \& Krassa, R.~F. 1971, \aap, 11, 218

\bibitem[{{Vallerga} {et~al.}(2004){Vallerga}, {Lallement}, {Lemoine},
  {Dalaudier}, \& {McMullin}}]{vallerga_etal:04a}
{Vallerga}, J., {Lallement}, R., {Lemoine}, M., {Dalaudier}, F., \& {McMullin},
  D. 2004, \aap, 426, 855, \dodoi{10.1051/0004-6361:20035887}

\bibitem[{{Witte}(2004)}]{witte:04}
{Witte}, M. 2004, \aap, 426, 835, \dodoi{10.1051/0004-6361:20035956}

\bibitem[{Witte {et~al.}(1993)Witte, Banaszkiewicz, \&
  Rosenbauer}]{witte_etal:93}
Witte, M., Banaszkiewicz, M., \& Rosenbauer, H. 1993, \asr, 13, (6)121

\bibitem[{{Witte} {et~al.}(1992){Witte}, {Rosenbauer}, {Keppler}, {Fahr},
  {Hemmerich}, {Lauche}, {Loidl}, \& {Zwick}}]{witte_etal:92a}
{Witte}, M., {Rosenbauer}, H., {Keppler}, E., {et~al.} 1992, \aaps, 92, 333

\bibitem[{{Wood} {et~al.}(2019){Wood}, {M{\"u}ller}, \&
  {M{\"o}bius}}]{wood_etal:19a}
{Wood}, B.~E., {M{\"u}ller}, H.-R., \& {M{\"o}bius}, E. 2019, \apj, 881, 55,
  \dodoi{10.3847.1538-4357/1538-4357ab2e74}

\bibitem[{{Wood} {et~al.}(2015){Wood}, {M{\"u}ller}, \&
  {Witte}}]{wood_etal:15a}
{Wood}, B.~E., {M{\"u}ller}, H.-R., \& {Witte}, M. 2015, \apj, 801, 62,
  \dodoi{10.1088/0004-637X/801/1/62}

\bibitem[{Wurz {et~al.}(2006)Wurz, Scheer, \& Wieser}]{wurz_etal:06a}
Wurz, P., Scheer, J., \& Wieser, M. 2006, {e-J. of Surface Science and
  Nanotechnology}, 4, 394

\bibitem[{{Zirnstein} {et~al.}(2015){Zirnstein}, {Heerikhuisen}, \&
  {McComas}}]{zirnstein_etal:15a}
{Zirnstein}, E.~J., {Heerikhuisen}, J., \& {McComas}, D.~J. 2015, \apjl, 804,
  L22, \dodoi{10.1088/2041-8205/804/1/L22}

\end{thebibliography}

\end{document}